\documentclass[applsci,article,accept,moreauthors,pdftex]{Definitions/mdpi}

\firstpage{1} 
\pubvolume{10}
\issuenum{23}
\articlenumber{8483}
\pubyear{2020}
\copyrightyear{2020}
\history{Received: 3 November 2020; Accepted: 24 November 2020; Published: 27 November 2020}

\usepackage{tikz}
\usetikzlibrary{fadings}
\usepackage[labelformat=simple]{subcaption}
\usepackage{epstopdf}
\DeclareCaptionLabelFormat{subcaptionlabel}{\normalfont(\textbf{#2}\normalfont)}
\usepackage{adjustbox}
\usepackage{siunitx}
\usepackage{stmaryrd}

\Title{Design and Optimization of a High-Time-Resolution Magnetic Plasma Analyzer (MPA)}

\Author{Benjamin Criton %
 $^{1,}$*$^{,\dagger}$\orcidB{}, Georgios Nicolaou $^{1,2}$\orcidA{} and Daniel Verscharen $^{1,3}$\orcidC{}}

\AuthorNames{Benjamin Criton, Georgios Nicolaou and Daniel Verscharen}

\address{%
$^{1}$ \quad Department of Space and Climate Physics, Mullard Space Science Laboratory, University College London, Dorking, Surrey RH5 6NT, UK; g.nicolaou@ucl.ac.uk (G.N.); d.verscharen@ucl.ac.uk (D.V.)\\
$^{2}$ \quad Southwest Research Institute, San Antonio, TX 78238, USA\\
$^{3}$ \quad Space Science Center, University of New Hampshire, Durham, NH 03824, USA \\}

\corres{Correspondence: benjamin.criton.19@ucl.ac.uk}

\firstnote{Current address: IRFU, CEA, Université Paris-Saclay, 91191 Gif-sur-Yvette, France.} 

\featuredapplication{This instrument concept is designed to analyze plasma onboard scientific spacecraft. Its measurements allow the construction of the velocity distribution functions of protons and $\alpha$-particles with high cadence to understand the small-scale kinetic processes in space~plasmas.}

\abstract{In-situ measurements of space plasma throughout the solar system require high time resolution to understand the plasma's kinetic fine structure and evolution. In this context, research~is conducted to design instruments with the capability to acquire the plasma velocity distribution and its moments with high cadence. We study a new instrument design, using a constant magnetic field generated by two permanent magnets, to analyze solar wind protons and $\alpha$-particles with high time resolution. We determine the optimal configuration of the instrument in terms of aperture size, sensor position, pixel size and magnetic field strength. We conduct this analysis based on analytical calculations and SIMION simulations of the particle trajectories in our instrument. We evaluate the velocity resolution of the instrument as well as Poisson errors associated with finite counting statistics. Our instrument is able to resolve Maxwellian and $\kappa$-distributions for both protons and $\alpha$-particles. This method retrieves measurements of the moments (density, bulk speed and temperature) with a relative error below 1\%. Our instrument design achieves these results with an acquisition time of only 5 ms, significantly faster than state-of-the-art electrostatic analyzers. Although the instrument only acquires one-dimensional cuts of the distribution function in velocity space, the simplicity and reliability of the presented instrument concept are two key advantages of our new design.}


\begin{document}

\section{Introduction}
\vspace{-6pt}

\subsection{Scientific Objectives}

The solar wind is a plasma flow, emitted by the Sun, that fills the entire heliosphere. As with other space plasmas, its properties have been studied extensively by spacecraft carrying in-situ plasma instruments \cite{WIND, Cluster, SWAP}. The kinetic state of such a plasma is fully described by the velocity distribution functions (VDFs) of the particles. Due to the very low collisionality in many space plasmas, thermal~particles often exhibit different distributions from a simple Maxwellian. For example,  $\kappa$-distributions are often used to describe the superthermal tails of the measured VDF \cite{kinetic_model_with_kappa, kappa_distribution, kappa_space_science, kappa_distri_book}. Thus, any~space-plasma instrument must resolve non-thermal properties of the particle VDFs. Moreover, as~the solar wind passes over the measurement point at a supersonic speed (approximately in the range from 300 to 800\,km/s) in the spacecraft frame, the acquisition of VDF variations must be achieved with high cadence. Assuming average solar-wind conditions, we translate this cadence requirement into a sampling time requirement for the instrument. Since many important kinetic processes occur on scales of the order of the proton gyro-radius $\rho_p$, we require that a suitable instrument for the study of ion-kinetic processes resolves spatial scales of size $\rho_p$. For a typical solar wind speed of $u=500$\,\text{km/s}, the resolution of a structure of size the proton gyro-radius at a heliocentric distance of 1\,au \mbox{($\rho_p=80\,\text{km}$)~\cite{proton_heating,Verscharen}} requires a sampling time of $T_s=2\pi\,\rho_p/2u\approx 500\,\text{ms}$ according to the Nyquist criterion. Modern space plasma instrumentation aims at sampling the VDF with a high enough cadence to study sub-ion-scale variations in the VDF. This is the direction taken by, for example, the~Debye mission \cite{Debye} and the THOR mission \cite{THOR}. The goal of our instrument design is a cadence of 5\,ms, which~is two orders of magnitude faster than the requirement to resolve structures with a size of the average $\rho_p$ at a heliocentric distance of 1\,au.

\subsection{State-of-the-Art Space Plasma Instruments}

Traditionally, most instruments measuring charged particles in the solar wind fall within one of two families: Faraday cups and electrostatic analyzers (ESAs). The Parker Solar Probe SWEAP instrument suite includes a Faraday cup (SPC) \cite{SWEAP, SWEAP_SPC} that achieves measurements with a cadence of 16\,Hz for a full distribution and greater than 128\,Hz in high-cadence mode (only one energy/charge window near the peak of the distribution is measured). The instrument has a 28$^\circ$ field of view (FOV) with 1$^\circ$ target resolution and achieves radial 1D measurements of the VDF. ESAs are often capable of performing 3D measurements of the distribution. The typical acquisition time of a full 3D VDF for an ESA is higher than the typical acquisition time of an 1D VDF for a Faraday cup. For instance, the~Hot Ion Analyzer (HIA) on board Cluster samples a full 3D VDF in 4\,s, while the Proton Alpha Sensor (PAS) on board Solar Orbiter samples a full 3D VDF in 1 s \cite{PAS_SO, PAS_nicolaou}. These longer acquisition times are a consequence of the measurement cycle in these instruments. Our instrument concept aims instead at high energy and time resolutions at the expense of a full 3D coverage. It is based on the successfully flown Magnetic Electron Ion Spectrometer (MagEIS) instrument onboard the Van Allen Probes, which~was designed for the measurement of energetic electrons in the Earth's radiation belts~\cite{MagEIS}. In the MagEIS design, the magnetic field created by two permanent magnets deflects the incoming electrons toward a set of silicon detectors. The hit positions of the electrons on the detector area are related to their energy according to the definition of the relativistic gyro-radius. The magnetic field is used to select energy bands for the energy-sensitive silicon detectors placed in the focal plane~\cite{MagEIS}. Our instrument design uses the same operation principle, although it relies on the measurement of the curvature radius by a position-sensitive sensor as its sole determination of the particle energies, which is sufficient for non-relativistic particles such as thermal ions in most space-plasma applications.

\subsection{Instrument Working Principle and Expectations}

Our instrument concept, the Magnetic Plasma Analyzer (MPA), comprises of two planar permanent magnets placed parallel to each other (face to face), creating a quasi-uniform magnetic field between them. The charged plasma particles follow different trajectories through the magnetic field depending on their energy. A position-sensitive detector is mounted on the side of the magnetic chamber. As a result, particles with different energies are detected at different positions on the detector. MPA detects all particles from a given look direction in the same acquisition step, independent of their energies. Indeed, the expected counts are independent of any voltage modulation (unlike ESAs or Faraday cups). Instead, the number of counts only depends on the physical parameters of the instrument and sensor to first order: aperture size, magnetic field strength and pixel size. This is the key advantage of MPA's working principle: the simultaneous measurement of all velocities in one acquisition step leads to a higher time resolution than ESAs. For our instrument concept, we define a pixel as a set length range on the position-sensitive sensor that can be sampled with one given anode. Small pixel sizes allow for a larger number of energy channels (better resolution) but result in fewer counts on each pixel. In this paper, we discuss the optimal geometry of the MPA instrument to measure proton and $\alpha$-particle VDFs, the two most abundant ion species in the solar wind. We then evaluate the performance of the instrument assuming Maxwellian and $\kappa$-distributions of the incoming particles. We assume a perfectly uniform magnetic field inside the instrument and zero magnetic field outside the instrument (except in our SIMION simulations in Section \ref{sec:simion}).

\section{Instrument Design}
\vspace{-6pt}

\subsection{Instrument Geometry and Functionning}

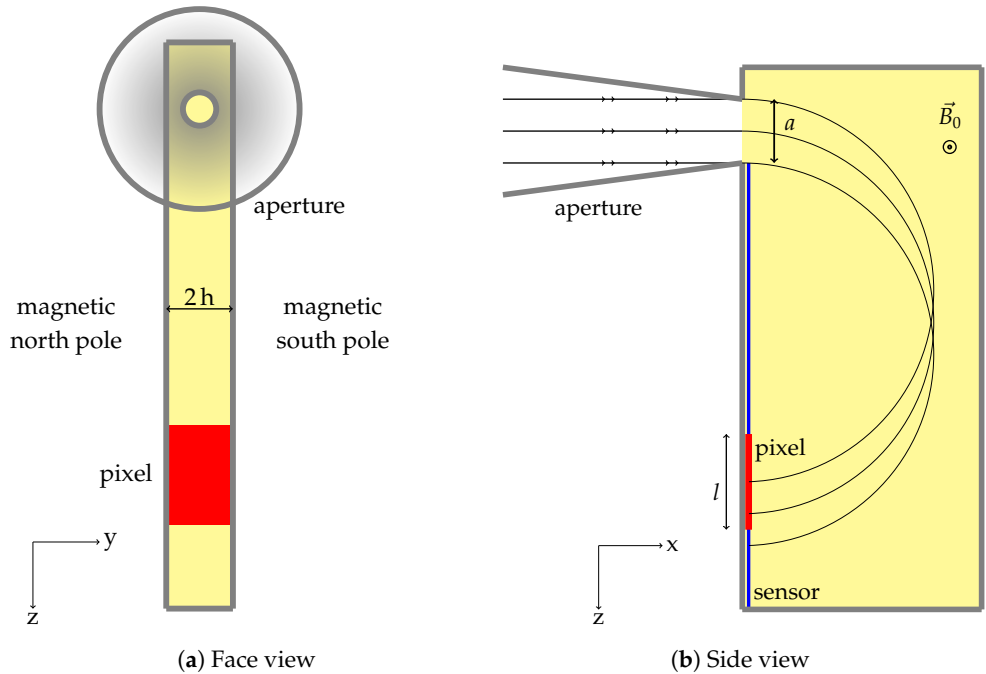
\begin{figure}[H]
\centering
\begin{minipage}[b]{0.4\textwidth}
\vspace*{1cm}
\resizebox{0.8\textwidth}{!}{
\begin{tikzpicture}
	\tikzfading[name=fade center,inner color=transparent!0,outer color=transparent!100]
	\fill [yellow, fill opacity=0.4] (-1,8) -- (1,8) -- (1,-9) -- (-1,-9);

	\fill [red, fill opacity=1] (-1,-3.5) -- (1,-3.5) -- (1,-6.5) -- (-1,-6.5);
	\node at (-2.2,-5) {\huge{pixel}};
	
	\node at (-4,0) {\huge{magnetic}};
	\node at (-4,-1) {\huge{north pole}};
	\node at (4,0) {\huge{magnetic}};
	\node at (4,-1) {\huge{south pole}};

	\draw[gray, line width=5pt] (-1,-9) -- (-1,8);
	\draw[gray, line width=5pt] (1,-9) -- (1,8);
	\draw[gray, line width=5pt] (-1,-9) -- (1,-9);
	\draw[gray, line width=5pt] (-1,8) -- (1,8);	
	\node at (3,3) {\huge{aperture}};
	\draw[gray, line width=5pt] (0,6) circle [radius=0.5];
	\draw[gray, line width=5pt] (0,6) circle [radius=3];	
	\path [draw=none,fill=gray, fill opacity = 0.9, path fading=fade center, even odd rule] (0,6) circle (0.5) (0,6) circle (3);
	\draw[black, very thick, <->] (-1,0) -- (1,0);
	\node[above] (0,0) {\huge{2\,h}};
	
	\draw[black, thick, ->] (-5,-7) -- (-5,-9);
	\node[below] at (-5,-9) {\huge{z}};
	\draw[black, thick, ->] (-5,-7) -- (-3,-7);
	\node[right] at (-3,-7) {\huge{y}};
		
\end{tikzpicture}}
\subcaption{Face view}
\label{schematic:face_view}
\end{minipage}
\begin{minipage}[b]{0.4\textwidth}
\vspace*{0.5cm}
\resizebox{\textwidth}{!}{
\begin{tikzpicture}
	\fill [yellow, fill opacity=0.4] (7.5,8) -- (15,8) -- (15,-9) -- (7.5,-9);
	\node at (14,6.5) {\huge{$\vec{B_0}$}};
	\draw[below, very thick] (14,5.5) circle [radius=0.2];
	\draw[below, ultra thick] (14,5.5) circle [radius=0.05];

	\draw[black, very thick] (0,5) -- (7.5,5);
	\draw[black, thick, ->] (5.4,5) -- (5.5,5);
	\draw[black, thick, ->] (5.1,5) -- (5.2,5);
	\draw[black, thick, ->] (3.4,5) -- (3.5,5);
	\draw[black, thick, ->] (3.1,5) -- (3.2,5);
	
	\draw[black, very thick] (0,6) -- (7.5,6);	
	\draw[black, thick, ->] (5.4,6) -- (5.5,6);
	\draw[black, thick, ->] (5.1,6) -- (5.2,6);
	\draw[black, thick, ->] (3.4,6) -- (3.5,6);
	\draw[black, thick, ->] (3.1,6) -- (3.2,6);
	
	\draw[black, very thick] (0,7) -- (7.5,7);	
	\draw[black, thick, ->] (5.4,7) -- (5.5,7);
	\draw[black, thick, ->] (5.1,7) -- (5.2,7);
	\draw[black, thick, ->] (3.4,7) -- (3.5,7);
	\draw[black, thick, ->] (3.1,7) -- (3.2,7);
	
	\draw[blue, line width=3pt] (7.7,5) -- (7.7,-9);
	\node at (8.9,-8.5) {\huge{sensor}};
	\draw[black, very thick, <->] (7,-3.5) -- (7,-6.5);
	\draw[red, line width=6pt] (7.7,-3.5) -- (7.7,-6.5);
	\node at (8.7,-4) {\huge{pixel}};
	\node at (6.7,-5.3) {\huge{$l$}};
	
	\draw[black, ultra thick, <->] (8.5,5) -- (8.5,7);
	\node at (9,6.2) {\huge{$a$}};
	\draw[gray, line width=5pt] (7.5,7) -- (0,8);
	\draw[gray, line width=5pt] (7.5,5) -- (0,4);
	\draw[gray, line width=5pt] (7.5,7) -- (7.5,8);
	\draw[gray, line width=5pt] (7.5,8) -- (15,8);
	\draw[gray, line width=5pt] (15,8) -- (15,-9);
	\draw[gray, line width=5pt] (15,-9) -- (7.5,-9);
	\draw[gray, line width=5pt] (7.5,-9) -- (7.5,5);
	\node at (3,3.5) {\huge{aperture}};
	
	\draw (7.5,6) arc[radius = 6, start angle= 90, end angle= -88];
	\draw (7.5,7) arc[radius = 6, start angle= 90, end angle= -88];
	\draw (7.5,5) arc[radius = 6, start angle= 90, end angle= -88];
	
	\draw[black, thick, ->] (3,-7) -- (3,-9);
	\node[below] at (3,-9) {\huge{z}};
	\draw[black, thick, ->] (3,-7) -- (5,-7);
	\node[right] at (5,-7) {\huge{x}};
	
\end{tikzpicture}}
\subcaption{Side view}
\label{schematic:side_view}
\end{minipage}
\caption{Basic geometry of MPA. The instrument model has a conical aperture of diameter $a$ at the entrance of the magnetic chamber (yellow area) with a semi angle of $2.5^\circ$. In Figure \ref{schematic:geometry}b, we represent the trajectories of three ions with equal mass, charge and speed but different offsets from the central aperture axis. ({\bf a}): Face view; ({\bf b}): Side view.}
\label{schematic:geometry}
\end{figure}

MPA consists of an aperture, a magnetic chamber and two permanent magnets, as illustrated in Figure \ref{schematic:geometry}. Ions within a limited field of view enter the magnetic chamber through the aperture. Inside~the gap between the two permanent magnets, ions are deflected by the magnetic field according to the Lorentz force:
\begin{equation}\label{equation:lorentz}
 m\,\frac{d\textbf{v}}{dt} = q\,\textbf{v} \wedge \textbf{B} \text{,}
\end{equation}
where $m$ and $q$ are the mass and charge of the ion, $\textbf{v}$ is its velocity vector and $\textbf{B}$ is the magnetic field inside the chamber. In a simplistic case in which $\textbf{B} = B_0\textbf{e}_y$ and the incoming particles have an initial velocity $\textbf{v}_0 = v_{0x}\textbf{e}_x$ with an initial position $\textbf{r}_0 = 0\,\textbf{e}_x+0\,\textbf{e}_y+0\,\textbf{e}_z$, the solution of Equation (\ref{equation:lorentz}) is:
\begin{equation}\label{equation:sol_motion}
\begin{cases}
	x(t) = \frac{v_{0x}}{\omega_0} \sin(\omega_0 t)\\
	z(t) = \frac{v_{0x}}{\omega_0} -\frac{v_{0x}}{\omega_0} \cos(\omega_0 t)
\end{cases}	\text{,}
\end{equation}
where $\omega_0 = qB_0/m$ is the homogeneous gyro-frequency inside the magnetic chamber. Equation (\ref{equation:sol_motion}) describes a circular trajectory of radius
\begin{equation}\label{equation:curvature_radius}
\tilde R = \frac{mv_{0x}}{qB_0} \text{.}
\end{equation}

Figure \ref{schematic:geometry}b illustrates the trajectories of three ions inside the chamber. According to Equation (\ref{equation:curvature_radius}), the curvature radius of the particles is directly proportional to their $v_{0x}$. Assuming that all particles are protons and knowing the magnetic field strength $B_0$, it is thus possible to measure the speed of the particles directly by measuring their curvature radius, represented by their hit distance in the sensor~plane.

\unskip

\subsection{Position of the Sensor to Obtain Optimal Velocity Resolution}\label{sec:sensor_position}

The position of the sensor within the magnetic chamber is crucial in order to obtain the optimal velocity resolution. We obtain the largest change of hit position with particle velocity when the sensor is perpendicular to the aperture (the sensor is mounted on the chamber's surface next to the aperture), as shown in Figure \ref{schematic:geometry}b. In this setup, the hit position $z$ is linked to the speed of a given particle by:
\begin{equation}\label{equation:hit_position}
z = 2\frac{mv_{0x}}{qB_0} \text{.}
\end{equation}

This expression is based on the assumption that the aperture is reduced to a point and that the particle velocity $\textbf{v}_0$ before entering the aperture is parallel to the $x$ axis. In Section \ref{section:instrument_performances}, we discuss the influence of the aperture size on the velocity determination. A second advantage of this sensor position, apart from having the highest velocity resolution, is the first-order focusing property which states that all particles with velocity $\textbf{v}_0$ and within a small solid-angle element are focused onto the same hit position in the sensor plane.
We demonstrate this property for a particle with $\textbf{v}_0 = v_{0x}\textbf{e}_x + v_{0z}\textbf{e}_z$ and
 $\textbf{r}_0 = \textbf{0}$. According to Equation (\ref{equation:lorentz}), we obtain:
\begin{equation}
\begin{cases}
\displaystyle
v_x(t) = v_{0x}\cos(\omega_0t)-v_{0z}\sin(\omega_0t) \\
\displaystyle
v_z(t) = v_{0z}\cos(\omega_0t)+v_{0x}\sin(\omega_0t) \\
\end{cases}	\text{,}
\end{equation}
and after integration:
\begin{equation}
\begin{cases}
\displaystyle
x(t) = x_0 + \frac{v_{0x}}{\omega_0}\sin(\omega_0t)+\frac{v_{0z}}{\omega_0}\cos(\omega_0t) \\
\displaystyle
z(t) = z_0 + \frac{v_{0z}}{\omega_0}\sin(\omega_0t)-\frac{v_{0x}}{\omega_0}\cos(\omega_0t) \\
\end{cases}	\text{.}
\end{equation}

Using $\cos(\alpha) = v_{0x}/\sqrt{v_{0x}^2+v_{0z}^2}$ and $\sin(\alpha) = v_{0z}/\sqrt{v_{0x}^2+v_{0z}^2}$, we obtain:
\begin{equation}
\begin{cases}
\displaystyle
x(t) = -\frac{\sqrt{v_{0x}^2+v_{0z}^2}}{\omega_0}\sin(\alpha) + \frac{\sqrt{v_{0x}^2+v_{0z}^2}}{\omega_0}\left[\cos(\alpha)\sin(\omega_0t)+\sin(\alpha)\cos(\omega_0t)\right] \\
\displaystyle
z(t) = \frac{\sqrt{v_{0x}^2+v_{0z}^2}}{\omega_0}\cos(\alpha) + \frac{\sqrt{v_{0x}^2+v_{0z}^2}}{\omega_0}\left[\sin(\alpha)\sin(\omega_0t)-\cos(\alpha)\cos(\omega_0t)\right] \\
\end{cases}	\text{,}
\end{equation}
leading to
\begin{equation}\label{equation:set_position}
\begin{cases}
\displaystyle
x(t) = -R\sin(\alpha) + R\sin\left(\omega_0t+\alpha\right) \\
\displaystyle
z(t) = R\cos(\alpha) - R\cos\left(\omega_0t+\alpha\right) \\
\end{cases}	\text{,}
\end{equation}
where we define the curvature radius as
\begin{equation}
R=\sqrt{v_{0x}^2+v_{0z}^2}/\omega_0	\text{.}
\end{equation}

The hit time $t_{\text{hit}}$ at which the particle hits the sensor plane fulfills the condition $x(t_{\text{hit}})= 0$. From~this condition, we find
\begin{equation}
t_{\text{hit}} = \frac{\pi-2\alpha}{\omega_0}
\end{equation}
resulting in
\begin{equation}\label{equation:z_hit}
z(t_{\text{hit}}) = 2\,R\,\cos(\alpha) = 2\,R\left[1-\frac{1}{2}\alpha^2 + o(\alpha^4)\right] \sim 2\,R	\text{.}
\end{equation}

Thus, for small incoming angles,  the hit position is not affected by the initial $z$-component of the particle velocity ($v_{0z}$). This approximation   is used in the remainder of the article to justify our assumption that all ions enter the instrument along its $x$-axis, despite potential (small) oblique entry angles of $\textbf{v}_0$.

\subsection{Dependence of the Field of View on Speed and Determination of the Magnetic Chamber Width}\label{sec:fov_dependence}

If the magnetic chamber is too narrow, fast particles with $v_{0y}\neq 0$ hit the magnet surfaces. These~particles would not be detected by the sensor or could potentially experience back-scattering as neutrals after charge-exchange, generating false counts in the sensor. This effect leads to an effective velocity-dependence of the field of view. For this characterization, we assume a particle inside the chamber with a trajectory according to Equation (\ref{equation:set_position}). Since the Lorentz force does not affect the trajectory of the particle along the parallel direction ($y$), the $y$-position of the particle after a time $t$ is $y(t) = v_{0y}\,t + y_0$, where $y_0$ accounts for a potential offset of the initial particle position. We define the half-width of the magnetic chamber as $h$. Without loss of generality, we assume that $v_{0y}\geq 0$. The~following arguments apply likewise to the case in which $v_{0y}<0$. A particle reaches the sensor plane without hitting a magnet surface if and only if
\begin{equation}
v_{0y}\,t + y_0 < h \,\,\forall \,t \,\in \left[0\,;\, t_{\text{hit}}\right] \text{ i.e., } \forall \,t \,\in\, \left[0\,;\,\frac{\pi-2\alpha}{\omega_0}\right]	\text{.}
\end{equation}

Thus, to fulfill the detection condition, the $y$-component of the initial velocity must follow this inequality: 
\begin{equation}
v_{0y} < \frac{\left(h-y_0\right)\,\omega_0}{\pi-2\alpha}	\text{,}
\end{equation}
which we re-write for small $\alpha$ as
\begin{equation}\label{equation:vy_ineq}
v_{0y} < \frac{\left(h-y_0\right)\,\omega_0}{\pi}	\text{.}
\end{equation}

Defining $\beta$ as the incoming angle of the particle in the $x$-$y$ plane, we find from Equation (\ref{equation:vy_ineq}) a relation between the out-of-plane angle and the measured velocity $v_{0x}$:
\begin{equation}
\beta = \arctan\left(\frac{v_{0y}}{v_{0x}}\right) < \arctan\left(\frac{q\left(h-y_0\right)\,B_0}{\pi\,m\,v_{0x}}\right)	\text{.}
\end{equation}

As a starting point, we choose to physically constrain the FOV of MPA to $5^{\circ}\times5^{\circ}$, setting a $5^\circ$ upper limit for the out-of-plane view angle of the instrument. This choice is based on the effective angular acceptance of the MagEIS Low and Medium instruments. Although the MagEIS Low and Medium chambers have a physical $20^\circ\times10^\circ$ FOV, their effective out-of-plane acceptance ranges from $10^\circ$ for the lowest to $3^\circ$ for the highest detectable energies \cite{MagEIS}. Considering the higher density of thermal ions measured by MPA compared to the highly energetic electrons measured by MagEIS, it is not necessary to use acceptance angles as large as $20^\circ$. Such a large acceptance angle would lead to a degradation of the focusing of ions in the sensor plane and production of more secondary emissions due to increased collisions of ions with the magnetic chamber walls. With our choice of a $5^\circ\times 5^\circ$ FOV, the out-of-plane view angle is given by
\begin{equation}
\gamma=\min\left[5^{\circ} , 2\times\arctan\left(\frac{q\left(h-y_0\right)\,B_0}{\pi\,m\,v_{0x}}\right)\right]	\text{.}
\end{equation}

\begin{figure}[H]
\centering
\includegraphics[scale=0.6]{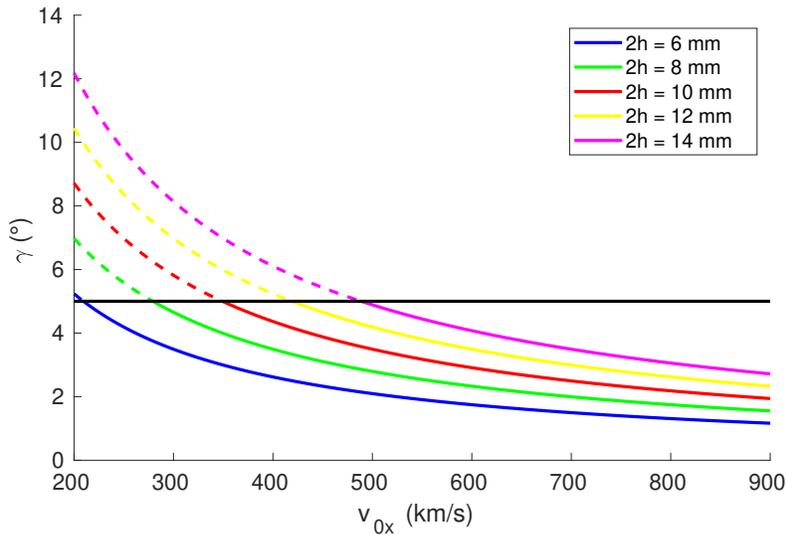}
\caption{Variation of the out-of-plane view angle $\gamma$ with velocity for different chamber width. These~curves simulate protons flying through the instrument chamber with different gap width (6, 8, 10, 12  and 14 mm), and a magnetic field strength of $B_0=0.1$\,T.}
\label{figure:field_of_view}
\end{figure}

Figure \ref{figure:field_of_view} shows the dependence of $\gamma$ on speed for different chamber widths when particles are assumed to enter the instrument at the center of the aperture: $y_0=0$. As expected, a larger gap between the magnets allows for a wider field of view at a given speed. MPA's effective out-of-plane view angle remains constant for particle speeds up to 500 km/s with a chamber gap of 14\,mm. Faster~particles will be detected with a narrower FOV. We note that the MagEIS instrument used a gap between its two magnets of approximately the same size (see Table \ref{table:MagEIS_parameters}).

To obtain a FOV of $5^\circ\times5^\circ$ on a wider range of velocities, two parameters can be adjusted: the chamber width or gap (as seen in Figure \ref{figure:field_of_view}) and the magnetic field strength. Wider magnetic chambers will induce higher levels of stray fields, as discussed by \citet{MagEIS}. On the other hand, we can keep the chamber narrow by increasing the magnetic field to the detriment of heavier magnetic materials. This trade-off must be conducted in further engineering work.

\subsection{From Counts to a VDF}

We assume a constant conical aperture with a half angle of 2.5$^\circ$ for the remainder of the article. Based on the experience with the MagEIS instrument, this aperture parameter guarantees a roughly constant viewing angle of the distribution function at a reasonable error on count numbers. This~aperture parameter is also the starting point of our discussion of the focusing property in Section~\ref{sec:sensor_position}. The aperture at the edge of the magnetic chamber is a disk of area $A$ and diameter $a$ (see~Figure~\ref{schematic:geometry}). The number of particles entering the conical aperture of area $A$ and semi-angle $\Theta_0/2 = 2.5^{\circ}$ during an acquisition time $\tau$ is
\begin{equation}
N = \tau\,A\int_0^{\infty}\int_0^{2\pi}\int_0^{\Theta_0/2}v\cos(\theta)\,f(v)\,v^2\sin(\theta)\,d\theta\,d\phi\,dv	\text{,}
\end{equation}
\begin{equation}
N = \tau\,A\,\frac{2\pi}{4}\,\big(1-\cos\left(\Theta_0\right)\big)\,\int_0^{\infty} v^3 f(v)\,dv	\text{,}
\end{equation}
where $f(v)$ is the (assumed to be) spherically symmetric VDF, which simplifies to:
\begin{equation}\label{equation:count_number_full}
N = G\,\int_0^{\infty} v^3 f(v)\,dv \text{,}
\end{equation}
where
\begin{equation}
G = \tau\,A\,\frac{\pi}{2}\,\left(1-\cos\left(\Theta_0\right)\right)
\end{equation}
is the ``geometric factor'' of the instrument. {The factor $G$ accounts for the key parameters of the instrument: aperture area, acquisition time and FOV.} Since $\Theta_0$ is small, we use
\begin{equation}
1-\cos\left(\Theta_0\right) \sim \frac{\Theta_0^2}{2}	\text{,}
\end{equation}
leading to
\begin{equation}\label{equation:geometric_factor}
G = \tau\,A\,\frac{\pi}{4}\,\Theta_0^2	\text{.}
\end{equation}

We note that the geometric factor derived here is not identical to the definition of the geometric factor of ESAs (see Equation (3) in \cite{bulk_parameters} or Equation (3.11) in \cite{JADE}) since $dE/E$ is not constant in the case of MPA. The position of the center of a detecting pixel $i$ in the position-sensitive sensor plane is {defined as}
\begin{equation}\label{equation:hit_distance}
z_i = \left(i-\frac{1}{2}\right)l+\frac{a}{2}	\text{\indent}\, i\in \llbracket\,1;n\,\rrbracket\,\text{,}
\end{equation}
where we place the origin of our coordinate system at the center of the aperture. According to Equation~(\ref{equation:hit_position}) and using the pixel setup shown in Figure \ref{schematic:geometry}b, the position of the center of pixel $i$ is linked to the velocity of a particle detected by this pixel as
\begin{equation}\label{equation:velocity_pixel}
v_i = \left[\left(i-\frac{1}{2}\right)l+\frac{a}{2}\right]\frac{qB_0}{2m} \text{,}
\end{equation}
where $v_i$ is the central velocity associated with pixel $i$ and $l$ is the width of the pixel. The difference of the speeds associated with two neighboring pixels is given by
\begin{equation}\label{equation:delta_v}
\delta v = v_{i+1}-v_i = \frac{qB_0l}{2m} \text{,}
\end{equation}
so that the number of counts (assuming a detection efficiency of 100\%) by pixel $i$ during an acquisition time $\tau$ is
\begin{equation}
C_i = G\,\int_{v_i - \frac{qBl}{4m}}^{v_i + \frac{qBl}{4m}} v^3 f(v)\,dv	\text{.}
\end{equation}

If we design the instrument so that the velocity step between two neighboring pixels is small compared to the measured velocities ($\delta v/v_0\sim 0$), we can re-write the number of counts in pixel $i$ for a given input distribution as
\begin{equation}\label{equation:expected_counts}
C_i = G\,\delta v\,v_i^3 f(v_i)	\text{.}
\end{equation}

The estimated distribution function obtained from the number of counts per pixel by the instrument is then simply
\begin{equation}\label{equation:estimated}
f_{est}(v_i) = \frac{C_i}{G\,\delta v\,v_i^3}	\text{.}
\end{equation}

This result is in agreement with the method used by \citet{determining_kappa} where the output VDF is obtained by inverting Equation (\ref{equation:expected_counts}). Equations (\ref{equation:expected_counts}) and (\ref{equation:estimated}) do not account for finite detection efficiencies and other non-ideal behaviors of the instrument. In general, the sensitivity of any type of particle detector depends on the speed, mass, charge and hit angle of the detected particles \cite{abs_detect_H, sensitivity_mcp}. We~introduce a finite, dimensionless efficiency parameter $\epsilon_{i_{\text{MCP}}} < 1$, where the subscript $i$ indicates pixel $i$ associated with speed $v_i$. In addition to the detector's characterization, the position-sensitive anode efficiency must be measured before launch \cite{Cluster}. We introduce a further finite, dimensionless~efficiency parameter $\epsilon_{i_{{anode}}}<1$ to correct for the anode efficiency. Moreover, the variation of the FOV with velocity formulated in Section \ref{sec:fov_dependence} must be included in the calculation of the geometric factor. We~introduce a finite, dimensionless normalization factor $g_i \leq 1$, which depends on velocity and is thus specific to each pixel $i$. Considering these effects, the corrected count number is given by
\begin{equation}\label{equation:corrected_counts}
\tilde C_i = \epsilon_{i_{\text{anode}}}\,\epsilon_{i_{\text{MCP}}}\,g_i\,C_i = \epsilon_{i_{\text{anode}}}\,\epsilon_{i_{\text{MCP}}}\,g_i\,G\,\delta v\,v_i^3 f(v_i)	\text{.}
\end{equation}

Lastly, we must apply a dead-time correction to the counting measurements. The true count number is obtained using the established formula \cite{evaluating_performances_analyzer, Cluster}
\begin{equation}\label{equation:true_counts}
C_{t_i} = \frac{\tilde C_i}{1-\frac{\tau_d}{\tau} \tilde C_i}	\text{,}
\end{equation}
where $\tau_d$ is the total effective dead time of the detector and the readout electronics    and   $\tau$ is the acquisition time. The parameters $\epsilon_{i_{\text{MCP}}}$, $\epsilon_{i_{{anode}}}$, $g_i$ and $\tau_d$ must be characterized and measured during the test and calibration phases. For the sake of simplicity in this conceptual study of MPA, we assume that $\epsilon_{i_{\text{MCP}}}=\epsilon_{i_{\text{anode}}}=g_i=1$, $\tau_d=0$,   and   use Equations (\ref{equation:expected_counts}) and (\ref{equation:estimated}) for our following simulations.

\subsection{Instrument Length}

As a minimum requirement, the $z$ dimension of the MPA's magnetic chamber must be longer than two curvature radii associated with the fastest particles. Similarly, the $x$ dimension must be deeper than one curvature radius (see Figure \ref{schematic:geometry}b). According to this requirement, the minimum dimensions of the instrument also depend on the magnetic-field strength, which defines the curvature radius. Keeping the remaining parameters constant, an increase of the magnetic-field strength increases the number of particles detected on each pixel, leading to a smaller statistical error in the measurement. However, it also increases the relative error in velocity measurements according to Equation (\ref{equation:delta_v}). The~MagEIS instrument provides us with a reasonable starting point for an achievable magnetic-field strength. The MagEIS properties are summarized in Table \ref{table:MagEIS_parameters}. We choose for MPA a starting value of 0.1\,T. In Section \ref{section:instrument_performances}, we study the influence of the magnetic field strength on the overall performance of our instrument.
\begin{table}[H]
\caption{Key parameters for MagEIS Low, Medium and High \cite{MagEIS,Van_allen_probe}. The mass of each instrument is estimated given the total mass of the suite containing four analyzers (one Low, one High and two Medium) \cite{Van_allen_probe}.}
\centering
\begin{tabular}{lccc}
\toprule
\textbf{Parameter}				&	\textbf{Low}					&	\textbf{Medium} 			&	 \textbf{High} 		\\
\midrule	
Energy range			&	20 keV--240 keV	&	80 keV--1200 keV	&	800 keV--4800 keV	\\
Magnetic field strength	&	0.055 T				&	0.16 T			&	0.48 T			\\
Magnetic chamber gap	&	7 mm				&	7 mm			& 	12 mm			\\
Field of view			&	$20^\circ\times 10^\circ$			&	$20^\circ\times 10^\circ$		&	$16^\circ\times19^\circ$	\\
Apperture geometry		&	2 mm $\times$ 5 mm			&	2 mm $\times$ 5 mm		&	10 mm $\times$ 5 mm	\\
Mass					&	8.5 kg				&	8.5 kg			&	8.5 kg			\\
\bottomrule
\end{tabular}
\label{table:MagEIS_parameters}
\end{table} 

In the solar wind, the proton bulk speed ranges from roughly 300   (slow solar wind) to 800~km/s (fast solar wind) (see Table 1 in  \cite{Marsch}, Figure 2 in   \cite{solar_wind_speed_sc} and \citet{Verscharen}). The speed also depends on latitude and heliospheric distance \cite{bulk_speed_solar_wind}. Using Equation (\ref{equation:hit_position}), we determine that the required length to measure protons with a speed of 800 km/s is 17 cm. However, the requirement to measure $\alpha$-particles as well makes necessary a longer instrument, since $\alpha$-particles have twice the charge of protons and four times their mass, leading to a larger hit distance in the sensor plane according to Equation (\ref{equation:hit_position}). Therefore, MPA must be twice as long as estimated in the proton-only case in order to measure $\alpha$-particles with the same speed as the protons. We choose for a first design a length of 30 cm. This~enables measurements of protons up to 1400 km/s and $\alpha$-particles up to 700 km/s. In contrast, the~MagEIS instrument is approximately 7 cm long, i.e. smaller,   and   thus lighter than our design.

\subsection{Summarized Instrument Geometry}

We summarize the relevant instrument parameters in Table \ref{table:instrument_characteristics}. We present a more detailed discussion of the aperture radius, pixel size and magnetic field in Section \ref{section:instrument_performances} to optimize MPA in terms of accuracy of the estimated VDF.
\begin{table}[H]
\caption{Summary  of the MPA characteristics.}
\centering
\begin{tabular}{lcc}
\toprule
\textbf{Parameter}	&&	\textbf{Value}	\\
\midrule
Aperture shape	&&	Cone of half-angle $2.5^\circ$	\\
Aperture entrance	&&	Disk of radius 1 mm	\\
Magnetic field	&&	Uniform at 0.1 T	\\
Instrument length	&&	30 cm	\\
Instrument height	&&	15 cm	\\
Instrument gap width	&&	14 mm	\\
Pixel size	&&	1 mm	\\
\bottomrule
\end{tabular}
\label{table:instrument_characteristics}
\end{table}
\unskip

\section{Instrument Performance}\label{section:instrument_performances}
\vspace{-6pt}

\subsection{Velocity Resolution}

The size of the aperture and of each pixel are two parameters that play key roles in the velocity resolution of the MPA instrument. {In addition, the detector sensitivity has an effect on the instrument performance. We discuss a trade-off between different detector technologies (micro-channel plates or multiple stacked channel electron multipliers) in Section \ref{sec:detector_tech}.} We calculate the velocity resolution of MPA in the case of a particle flying in a plane parallel to the magnets ($\textbf{v}_0 = v_{0x}\textbf{e}_x$). We determine the maximum velocity of a particle hitting pixel $i$ at the distance given by Equation (\ref{equation:hit_distance}) from the center of the aperture with a velocity $v_i = q\,B_0\,R_i/m$ expressed in Equation (\ref{equation:velocity_pixel}). The fastest particle that this pixel measures comes from the top of the aperture and hits the bottom of the pixel. Thus,
\begin{equation}
2\,R_{max} = 2\,R_i + \frac{a}{2} + \frac{l}{2} \text{,}
\end{equation}
which leads to a maximum speed associated with this pixel of
\begin{equation}
v_{max} = v_i + \frac{a+l}{4}\frac{qB_0}{m}\text{.}
\end{equation}

By analogy, the minimum speed associated with pixel $i$ is
\begin{equation}
v_{min} = v_i - \frac{a+l}{4}\frac{qB_0}{m}\text{.}
\end{equation}

Consequently, the velocity resolution of the instrument is
\begin{equation}
\Delta\,v = v_{max} - v_{min} = \frac{a+l}{2}\frac{qB_0}{m}	\text{.}
\end{equation}

\begin{figure}[H]
\centering
\scalebox{.85}[0.85]{\includegraphics[scale=0.5]{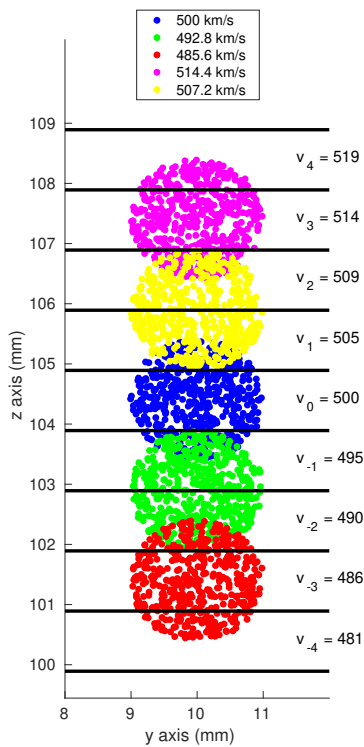}}
\caption{Five beams of particles are simulated using the SIMION software. Each beam has a set speed and a cylindrical (uniform) distribution of positions at the aperture of the instrument. The radius of the beam is 1\,mm (2-mm disk aperture). Pixels are the areas separated by black horizontal lines.}
\label{figure:hit_pos_beam}
\end{figure}

With MPA's characteristics presented in Table \ref{table:instrument_characteristics}, we obtain a resolution of 14.4 km/s, which yields for a 500 km/s solar wind a relative resolution of $\Delta\,v/v_0 = 2.8\%$. This phenomenon is illustrated in Figure \ref{figure:hit_pos_beam} where the hit positions of five beams of protons, separated by velocity steps of 7.2\,km/s, are~depicted. Pixels of 1 mm width are represented between the black lines. A pixel positioned to measure particles with speed $v_0$ can in fact detect particles with speed in $\left[v_0 - \Delta v/2\,;\,v_0 + \Delta v/2\right]$. Thus, the~uncertainty on the speed of one detected particle is $\Delta v$.

\unskip

\subsection{Errors Based on Counting Statistics}\label{sec:error_counting}

\begin{figure}[H]
\centering
\includegraphics[scale=0.5]{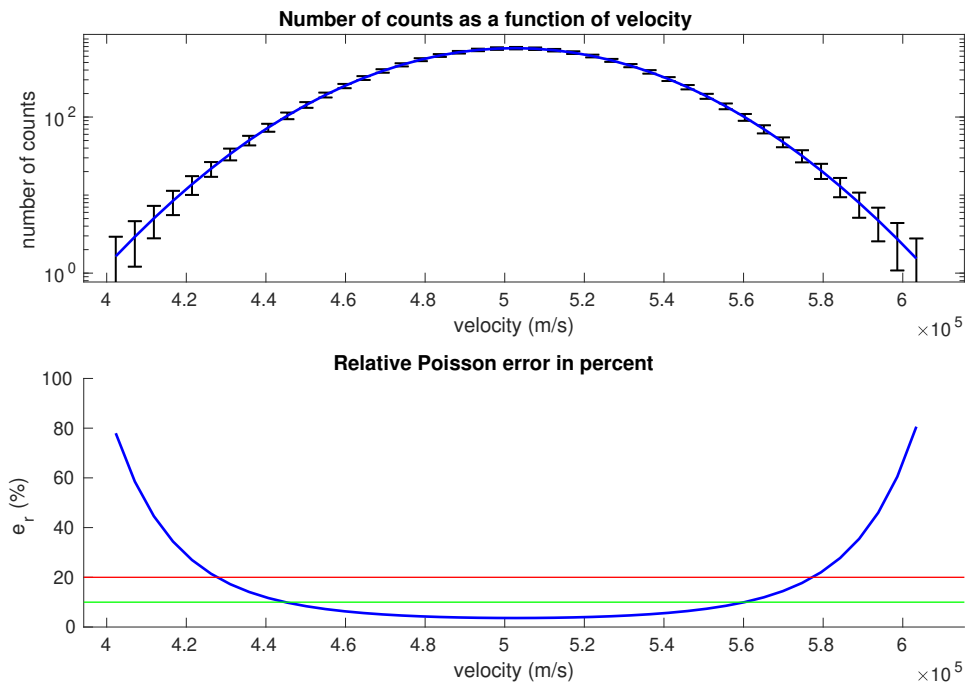}
\caption{Evolution of the counting errors with particle speed. The top plot shows the $2\sigma$ error bars on the expected counts for each pixel. The bottom plot represents the relative error ($e_r$) in percent made on the determination of the speed. The 10\% (green) and 20\% (red) levels are represented.}
\label{figure:poisson_errors}
\end{figure}

\begin{figure}[H]
\centering
\includegraphics[scale=0.7]{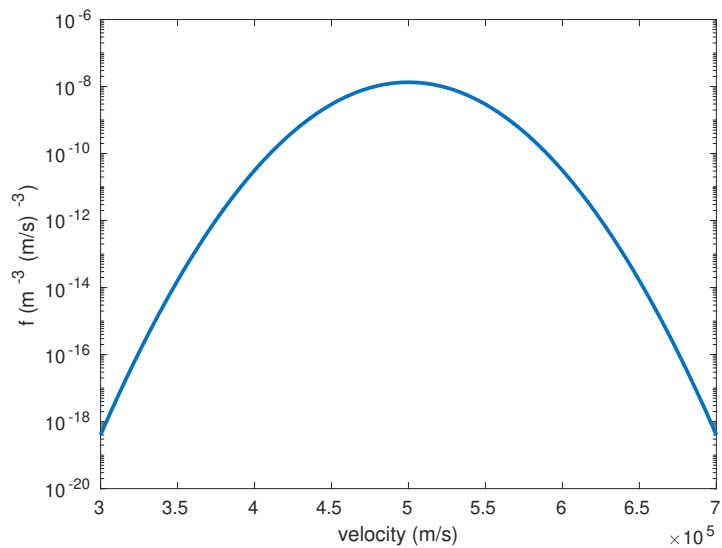}
\caption{Maxwellian VDF with density 5\,cm$^{-3}$, temperature 100,000\,K and bulk speed 500\,km/s. This~distribution function serves as the input for our calculations in Section \ref{sec:error_counting}.}
\label{figure:maxwellian_vdf}
\end{figure}

In addition to the velocity errors based on the finite resolution of the instrument, the detection of particles also introduces uncertainties due to the counting statistics. The number of counts $N_i$ that the sensor measures in pixel $i$ follows a Poisson distribution:
\begin{equation}
P(N_i) = e^{-C_i}\frac{C_i^{N_i}}{N_i!}	\text{,}
\end{equation}
where $C_i$ is the expected number of counts for velocity $v_i$ presented in Equation (\ref{equation:expected_counts}). The standard deviation of the Poisson distribution is $\sigma = \sqrt{C_i}$. The number of counts for each pixel with a confidence interval of $2\,\sigma$ is presented in Figure \ref{figure:poisson_errors}. We use an acquisition time of 5 ms with a disk aperture of 1\,mm radius and a pixel size of 1\,mm in width. The solar wind is modeled with a Maxwellian VDF with a density of 5\,cm$^{-3}$, {as represented in Figure \ref{figure:maxwellian_vdf}}. Due to the high number of counts per pixel, the~Poisson errors are low over a wide range of velocities around the center of the VDF. We obtain less than 10\% of relative error for speeds ranging from 440  to 560\,km/s, which corresponds to $\pm 60\,\mathrm{km/s}$ around the central velocity of the VDF. The Poisson errors are, however, linked to the size of the pixels. The larger the pixels are, the lower the relative Poisson errors are due to the larger $N_i$ for each pixel, \mbox{and vice versa}.

\unskip

\unskip

\subsection{Optimization of the Aperture Size, Pixel Size, and Magnetic Field Strength}

\begin{figure}[H]
	\centering
	\includegraphics[scale=0.7]{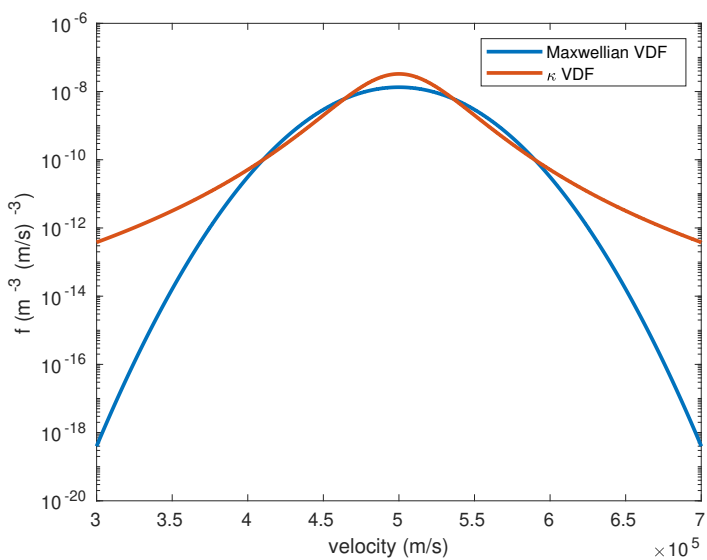}
	\caption{Maxwellian and $\kappa$ VDFs with parameters summarized in Table \ref{table:intermediate_wind}. These distribution functions serve as inputs for the following simulations.}
	\label{figure:input_vdf}
\end{figure}

We conduct test-particle simulations to quantify the relative errors made on VDF parameters for different aperture sizes and pixel sizes. We assume an idealized homogeneous magnetic field inside the instrument and use Equation (\ref{equation:hit_position}) to determine the hit position of the particles. We inject particles with a Maxwellian distribution and a $\kappa$-distribution starting with a uniform distribution of particle locations on the instrument's disk aperture. {We represent both input distributions (Maxwellian~and $\kappa$-distribution) in Figure \ref{figure:input_vdf}.} We obtain estimates of the VDF parameters such as number density ($n$), thermal speed ($v_T$) and bulk speed ($U$) by fitting, using a Levenberg--Marquardt algorithm, for~every set of aperture and pixel size. For the $\kappa$-distribution, we require a fourth parameter, the $\kappa$-index. We~run these simulations for different types of solar wind: slow, intermediate and fast. The instrument and solar wind parameters are summarized in Table \ref{table:intermediate_wind}. 

\unskip

\begin{table}[H]
\caption{Input values for our simulations.}
\centering
\begin{tabular}{lcc}
\toprule
\textbf{Parameter}	&&	\textbf{Value}	\\
\midrule
Density	&&	5 cm$^{-3}$	\\
Temperature	&&	100,000 K	\\
Bulk speed	&&	500 km/s	\\
$\kappa$ &&	3	\\
Measurement duration	&&	5 ms	\\
FOV	&&	$5^\circ\times5^\circ$	\\
Magnetic field strength	&&	0.1 T	\\
\bottomrule
\end{tabular}
\label{table:intermediate_wind}
\end{table}

Figure \ref{figure:intermediate_results} shows the relative error made on the determination of the plasma moments: density, thermal speed and on its bulk speed as a function of aperture diameter and pixel size for the particular case of an intermediate solar wind. Simulations for slow and fast winds yield similar results (not~shown here). Regions in parameter space for which the relative error on the determination of the VDF parameters is low appear in blue in Figure \ref{figure:intermediate_results}. From this analysis, we choose an ideal pixel size of 1\,mm and an ideal aperture diameter of 2\,mm. We conduct the same type of simulations to evaluate the impact of the magnetic field strength on the resolution of the instrument. We plot the relative errors of the VDF parameters as a function of magnetic field strength and pixel size. The results are shown in Figure \ref{figure:group_B}. According to this set of simulations, the maximum value of the magnetic field that still achieves low relative errors (below 10\%) is approximately 0.2\,T. Therefore, it is possible to divide  the length of MPA from our initial value of 30\,cm by a factor of two without major loss of accuracy.

\begin{figure}[H]
 \begin{minipage}[t]{.45\linewidth}
 \centering
 \includegraphics[width=\textwidth]{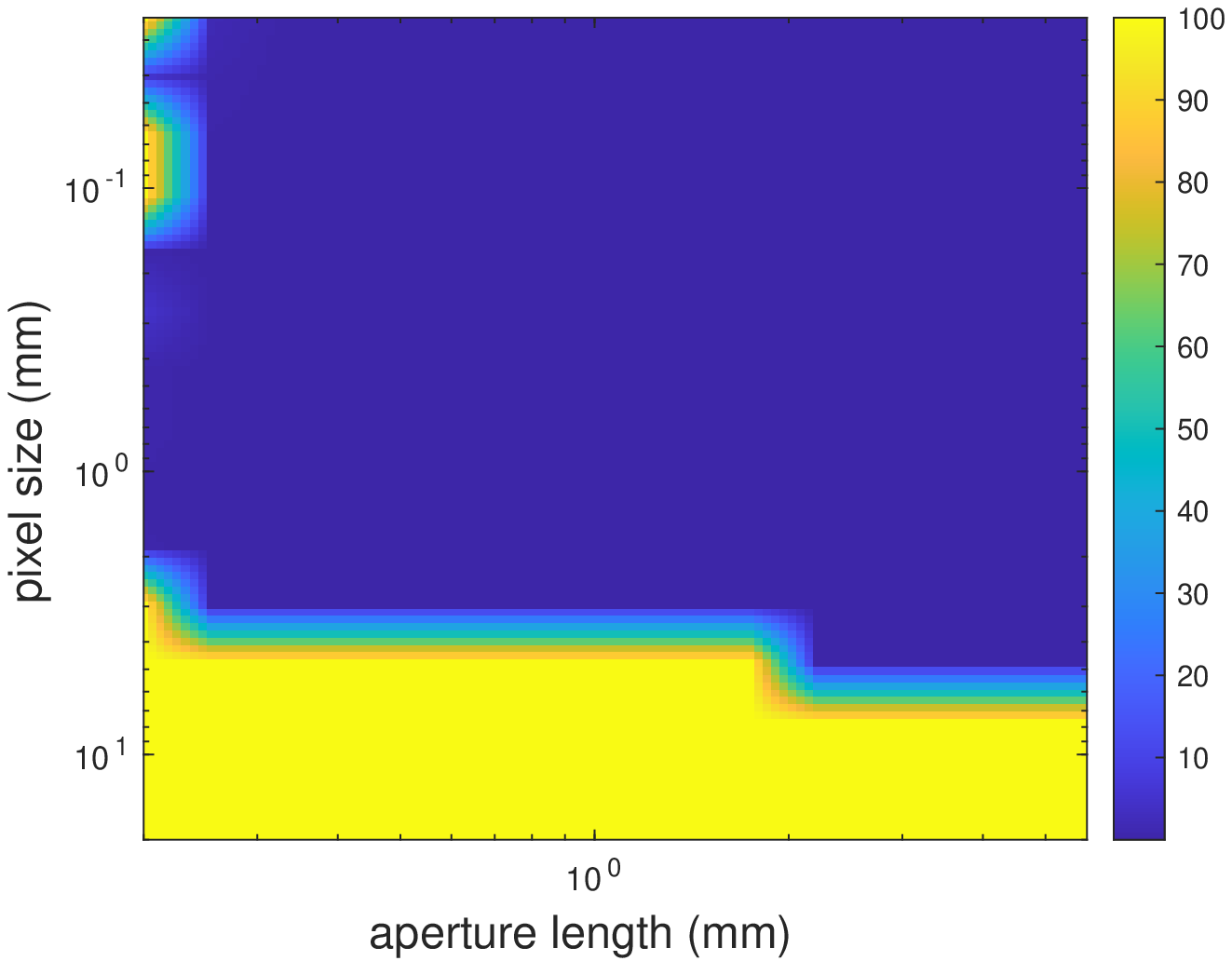}
 \subcaption{Bulk speed---Maxwellian distribution}
 \label{bulk_maxwellian_mean}
 \end{minipage}
 \hfill
 \vspace{0.2cm}
 \begin{minipage}[t]{.45\linewidth}
 \centering
 \includegraphics[width=\textwidth]{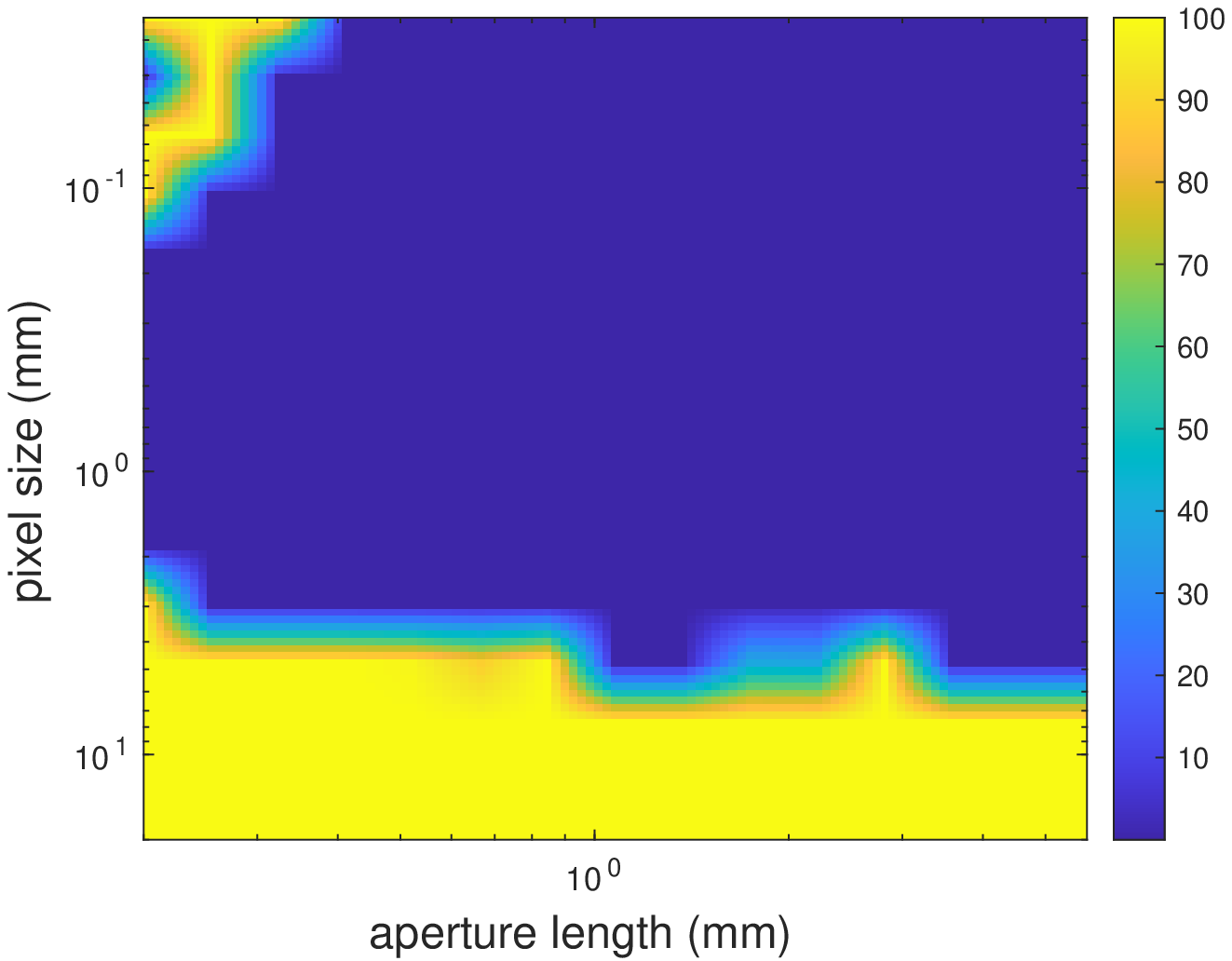}
 \subcaption{Bulk speed---$\kappa$-distribution}
 \label{bulk_kapa_mean}
 \end{minipage}

 \begin{minipage}[t]{.45\linewidth}
 \centering
 \includegraphics[width=\textwidth]{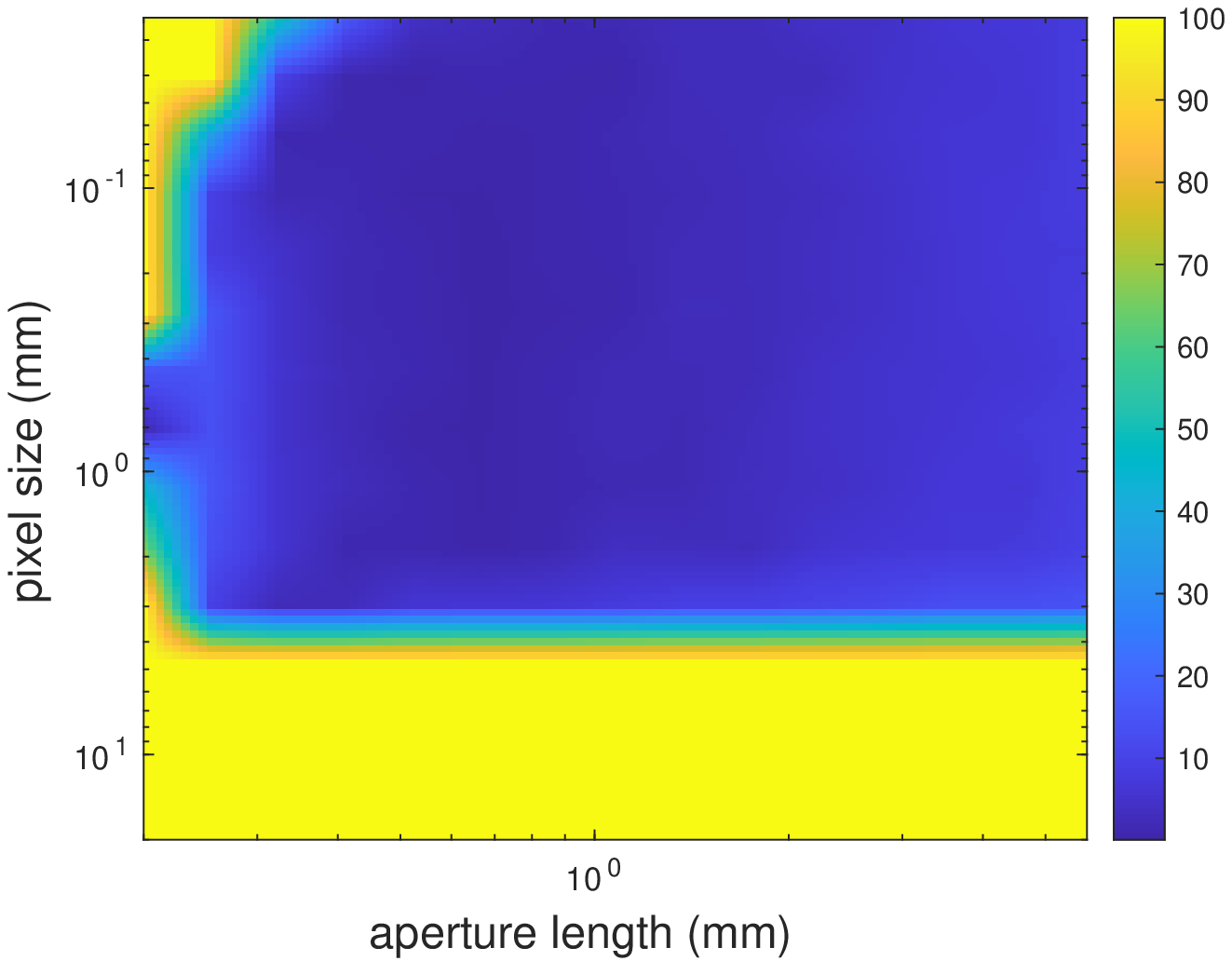}
 \subcaption{Density---Maxwellian distribution}
 \label{density_maxwellian_mean}
 \end{minipage}
 \hfill
 \vspace{0.2cm}
 \begin{minipage}[t]{.45\linewidth}
 \centering
 \includegraphics[width=\textwidth]{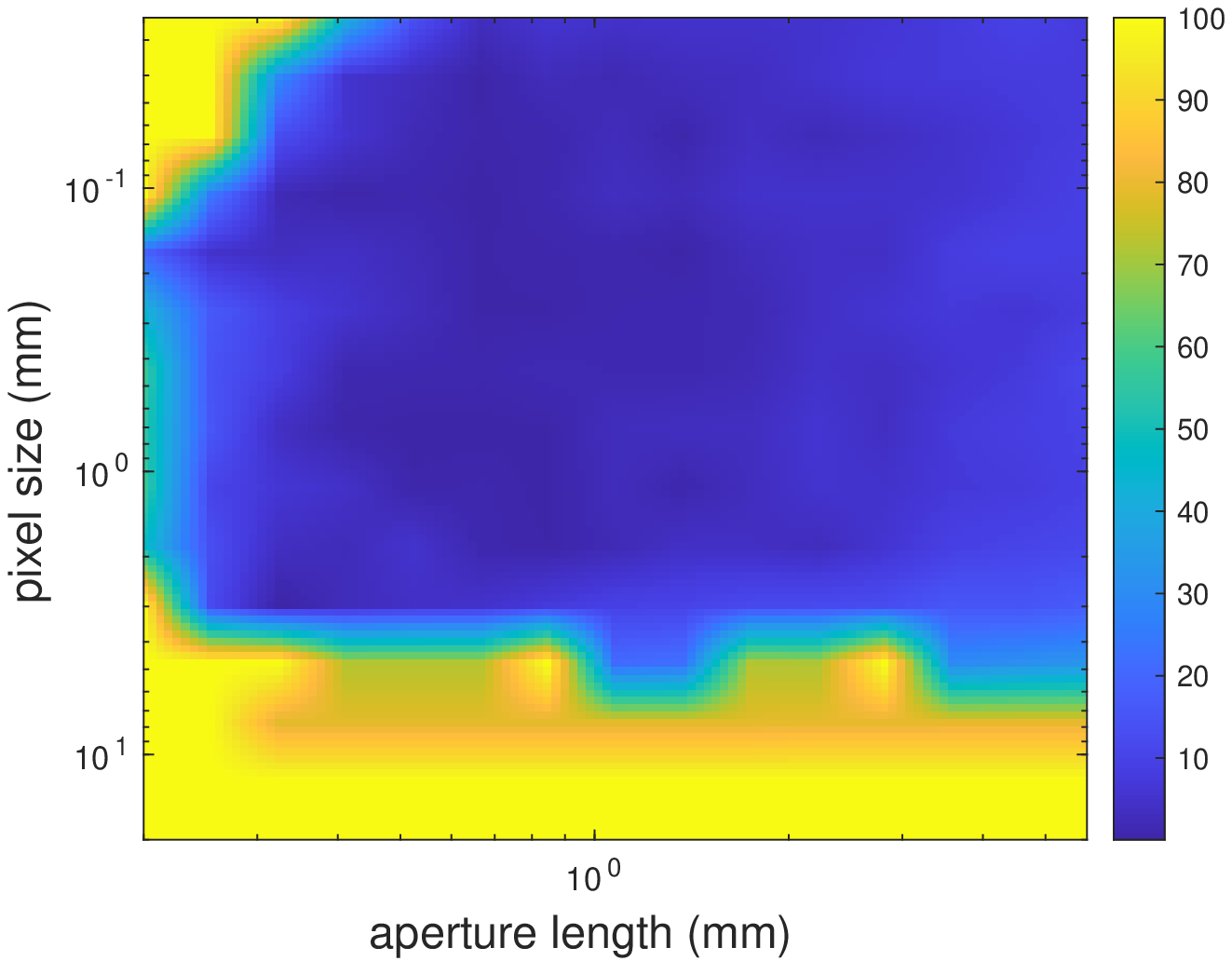}
 \subcaption{Density---$\kappa$-distribution}
 \label{density_kappa_mean}
 \end{minipage}

 \caption{\textit{Cont}.}
\end{figure}
 \begin{figure}[H]\ContinuedFloat
\centering
\setcounter{subfigure}{4}

 \begin{minipage}[t]{.45\linewidth}
 \centering
 \includegraphics[width=\textwidth]{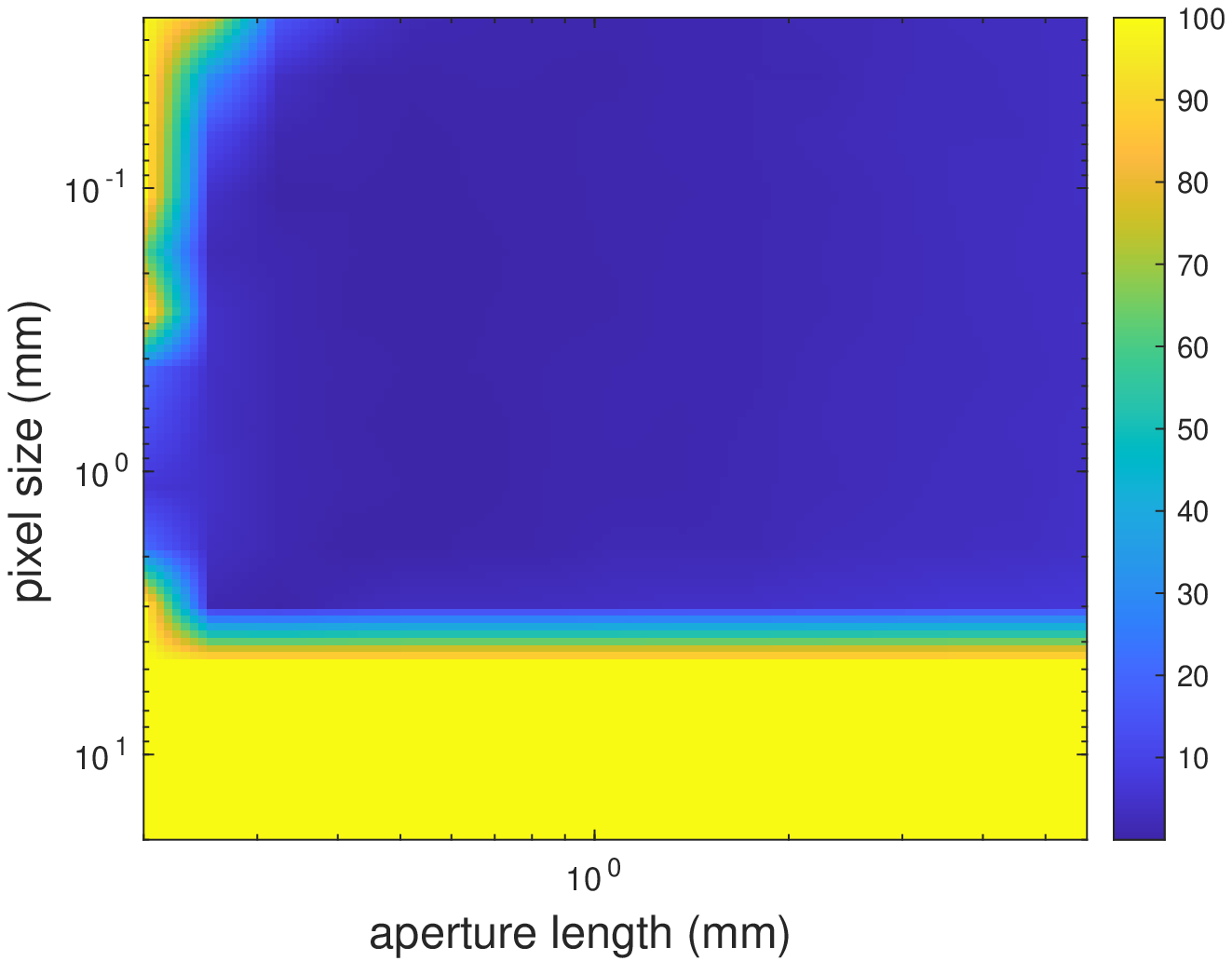}
 \subcaption{Thermal speed---Maxwellian distribution}
 \label{vt_maxwellian_mean}
 \end{minipage}
 \hfill
 \vspace{0.2cm}
 \begin{minipage}[t]{.45\linewidth}
 \centering
 \includegraphics[width=\textwidth]{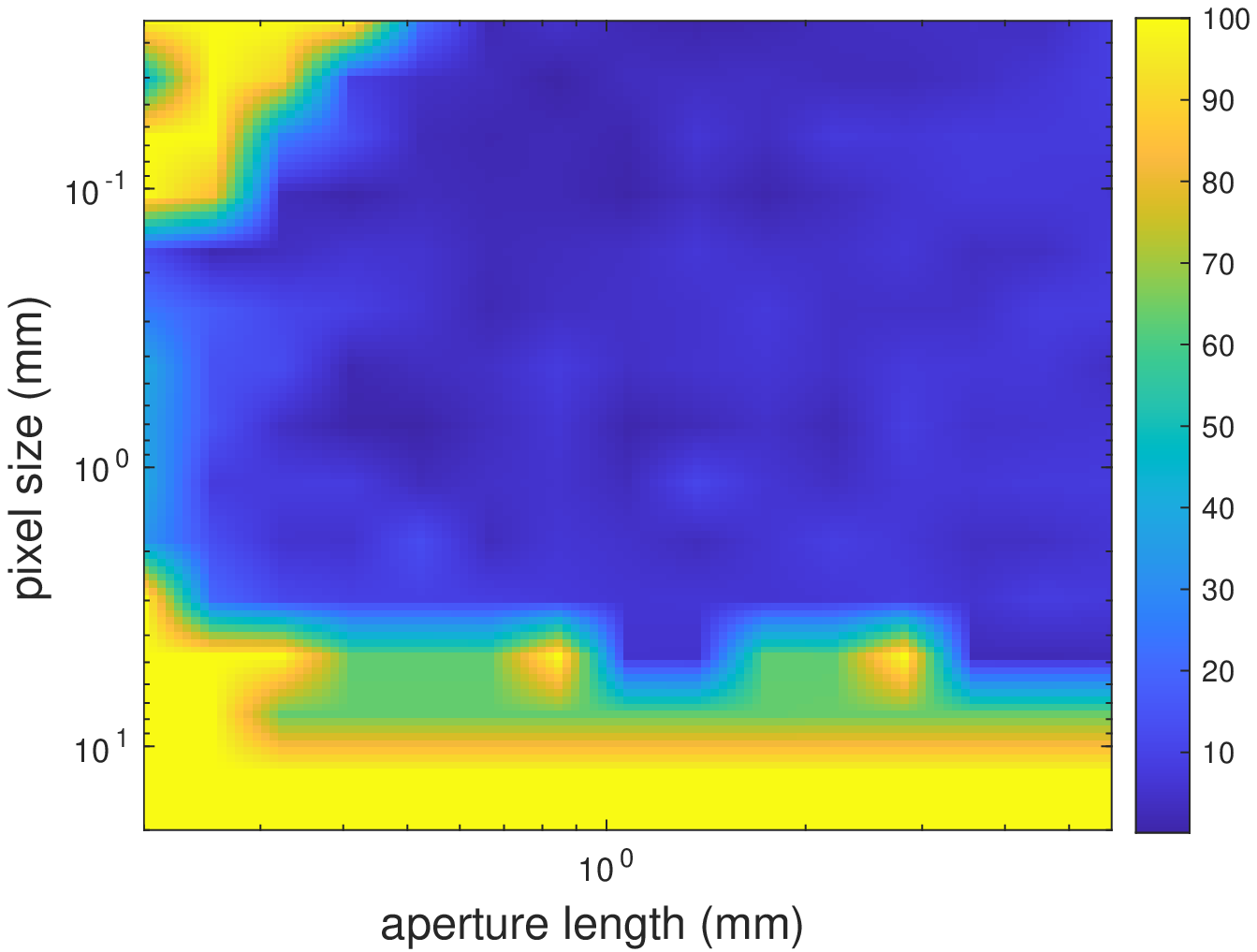}
 \subcaption{Thermal speed---$\kappa$-distribution}
 \label{vt_kappa_mean}
 \end{minipage} 
	\hspace*{0.2cm}
	\begin{minipage}[t]{\linewidth}
 \centering
 \includegraphics[width=0.45\textwidth]{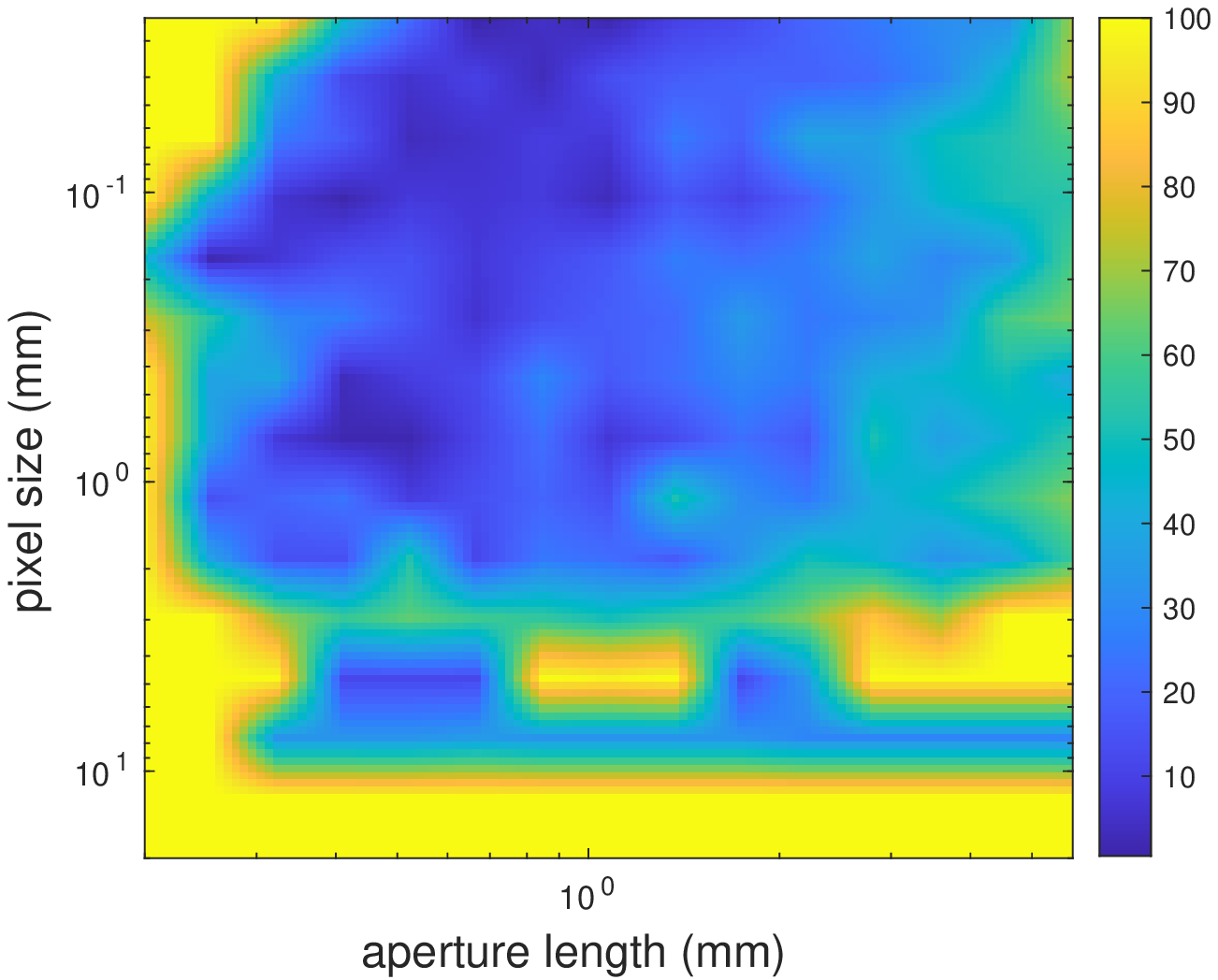}
 \subcaption{$\kappa$-parameter}
 \label{kappa_mean}
 \end{minipage}
 \vspace{6pt}
 \caption{Relative errors in percent made on measurements of an intermediate solar wind (bulk speed of 500 km/s, temperature of 100,000 K and density of 5 cm$^{-3}$) as a function of aperture and pixel size. Panels (\textbf{a}) through (\textbf{g}) show the resulting relative errors for all of the fit parameters of the Maxwellian and $\kappa$-distributions.}
 \label{figure:intermediate_results}
\end{figure}
\unskip

\begin{figure}[H]
 \begin{minipage}[t]{.45\linewidth}
 \centering
 \includegraphics[width=\textwidth]{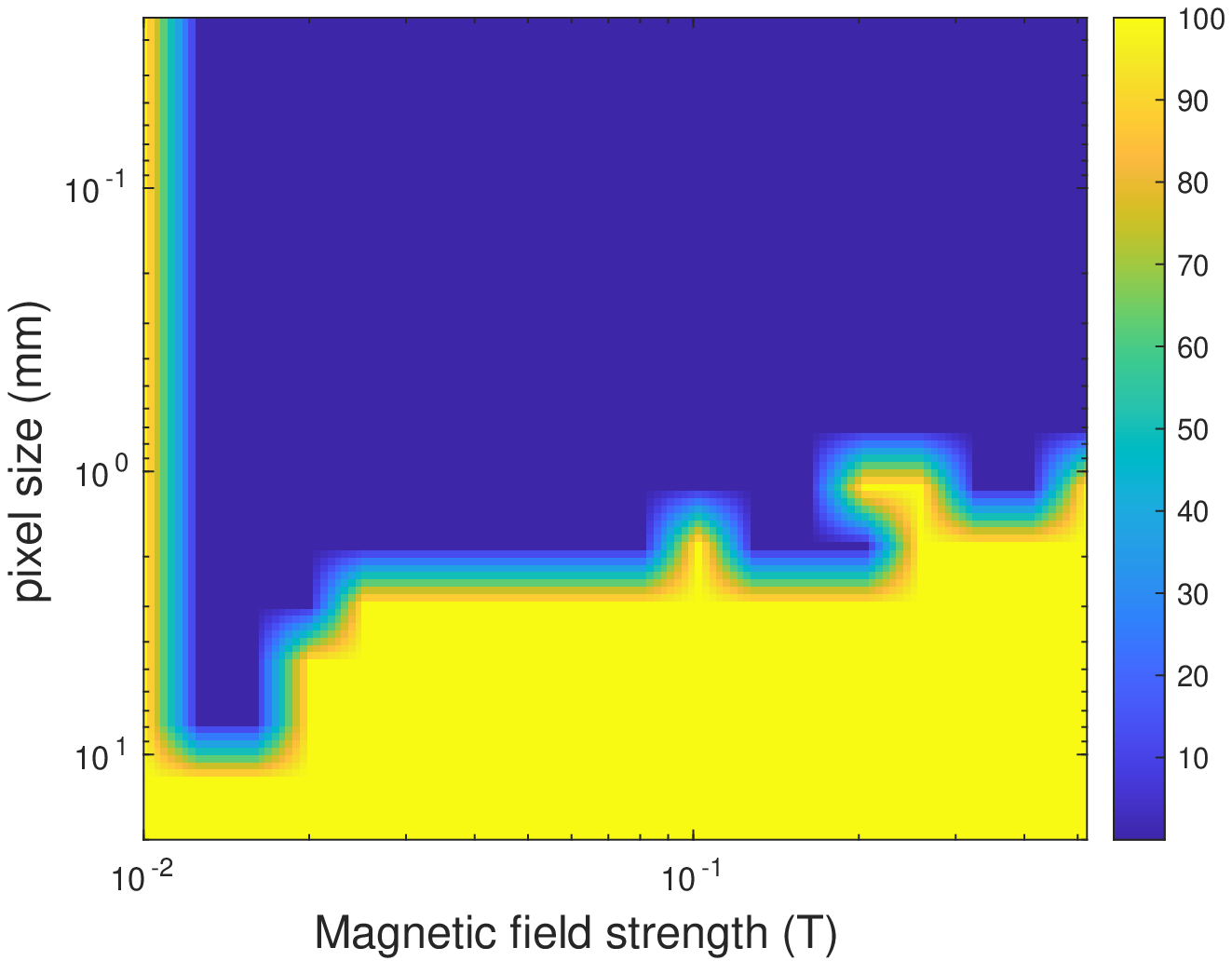}
 \subcaption{Bulk speed---Maxwellian distribution}
 \label{bulk_maxwellian_B}
 \end{minipage}
 \hfill
 \vspace{0.2cm}
 \begin{minipage}[t]{.45\linewidth}
 \centering
 \includegraphics[width=\textwidth]{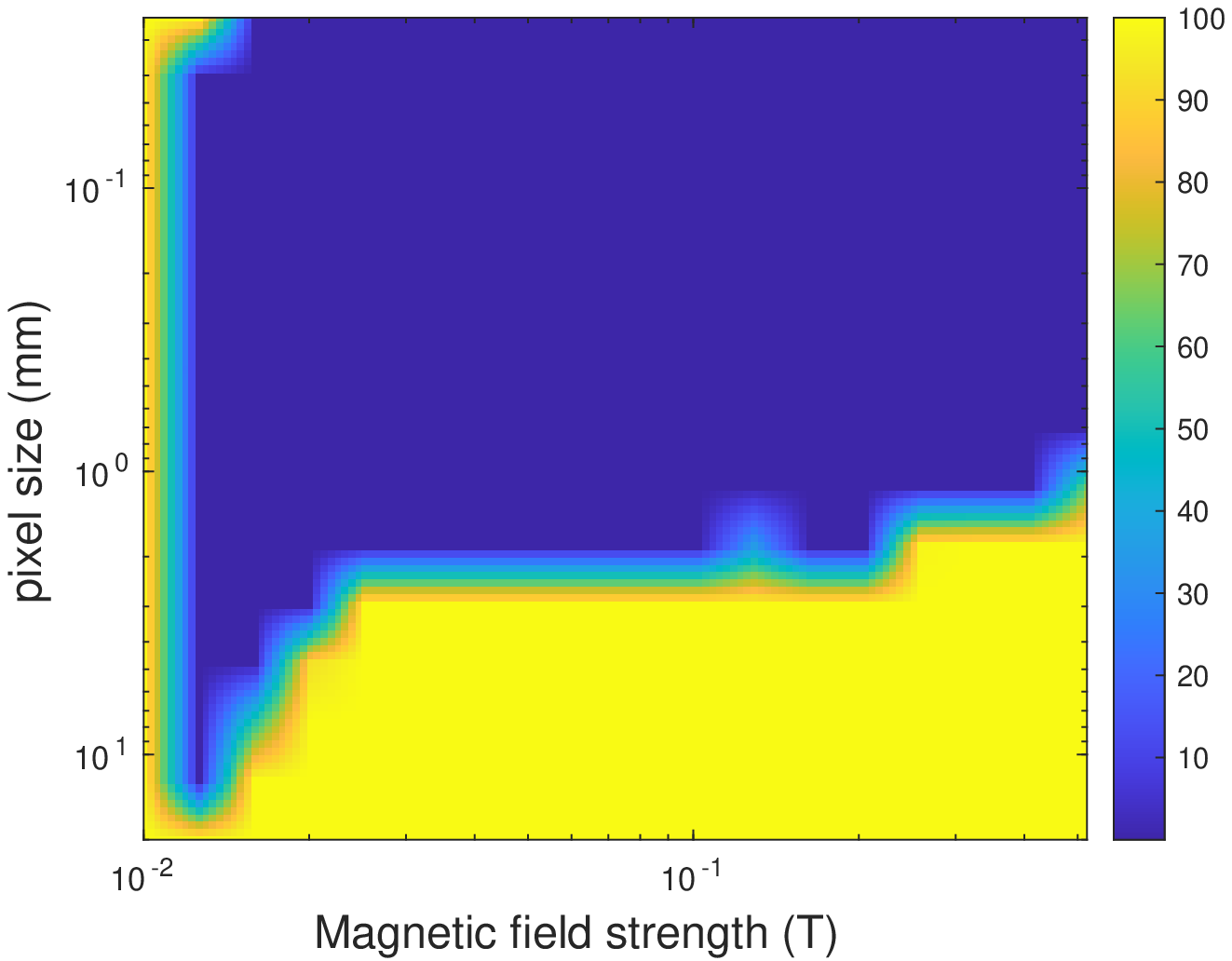}
 \subcaption{Bulk speed---$\kappa$-distribution}
 \label{bulk_kappa_B}
 \end{minipage}
 
 \caption{\textit{Cont}.}
\end{figure}
 \begin{figure}[H]\ContinuedFloat
\centering
\setcounter{subfigure}{2}

 \begin{minipage}[t]{.45\linewidth}
 \centering
 \includegraphics[width=\textwidth]{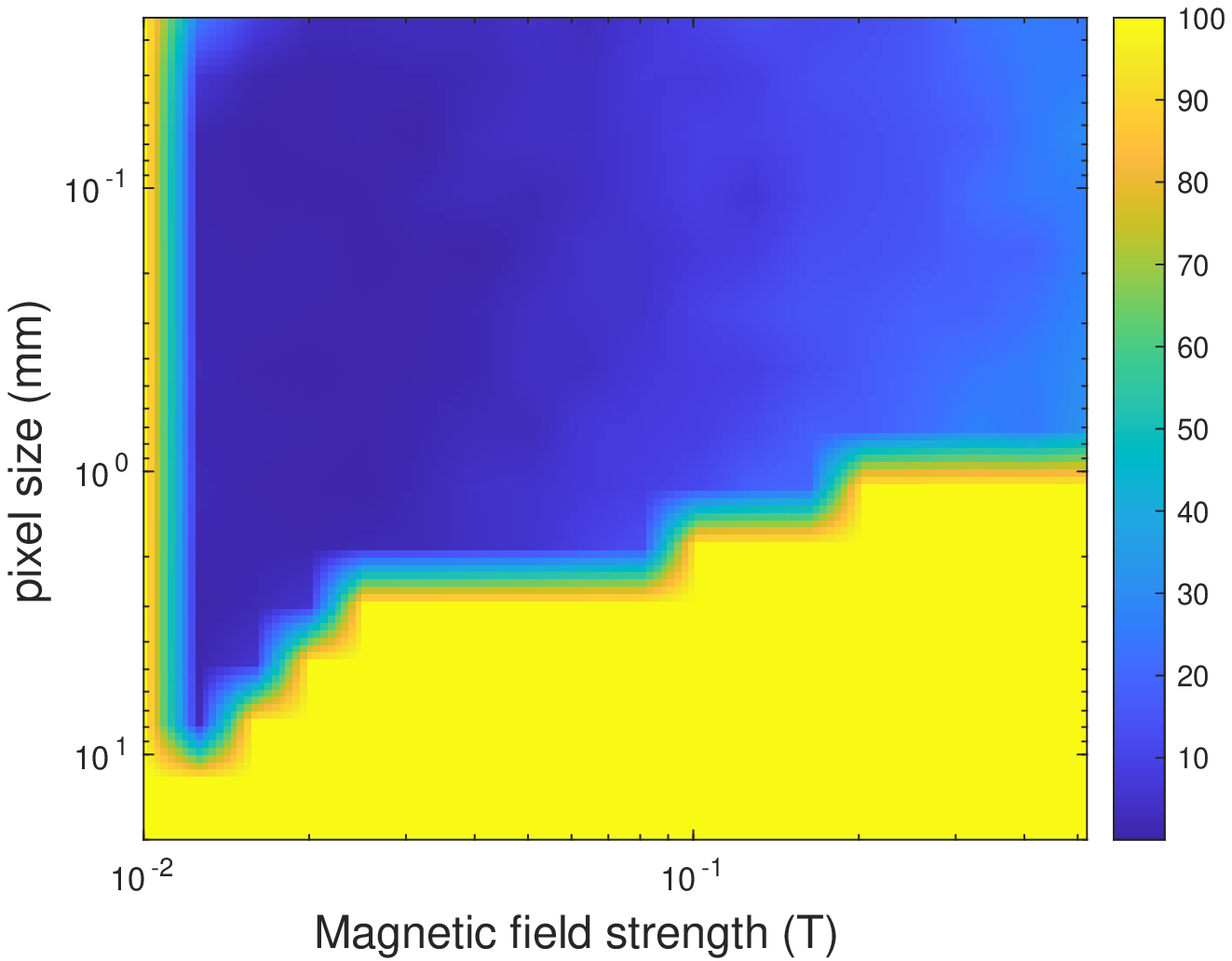}
 \subcaption{Density---Maxwellian distribution}
 \label{density_maxwellian_B}
 \end{minipage}
 \hfill
 \vspace{0.2cm}
 \begin{minipage}[t]{.45\linewidth}
 \centering
 \includegraphics[width=\textwidth]{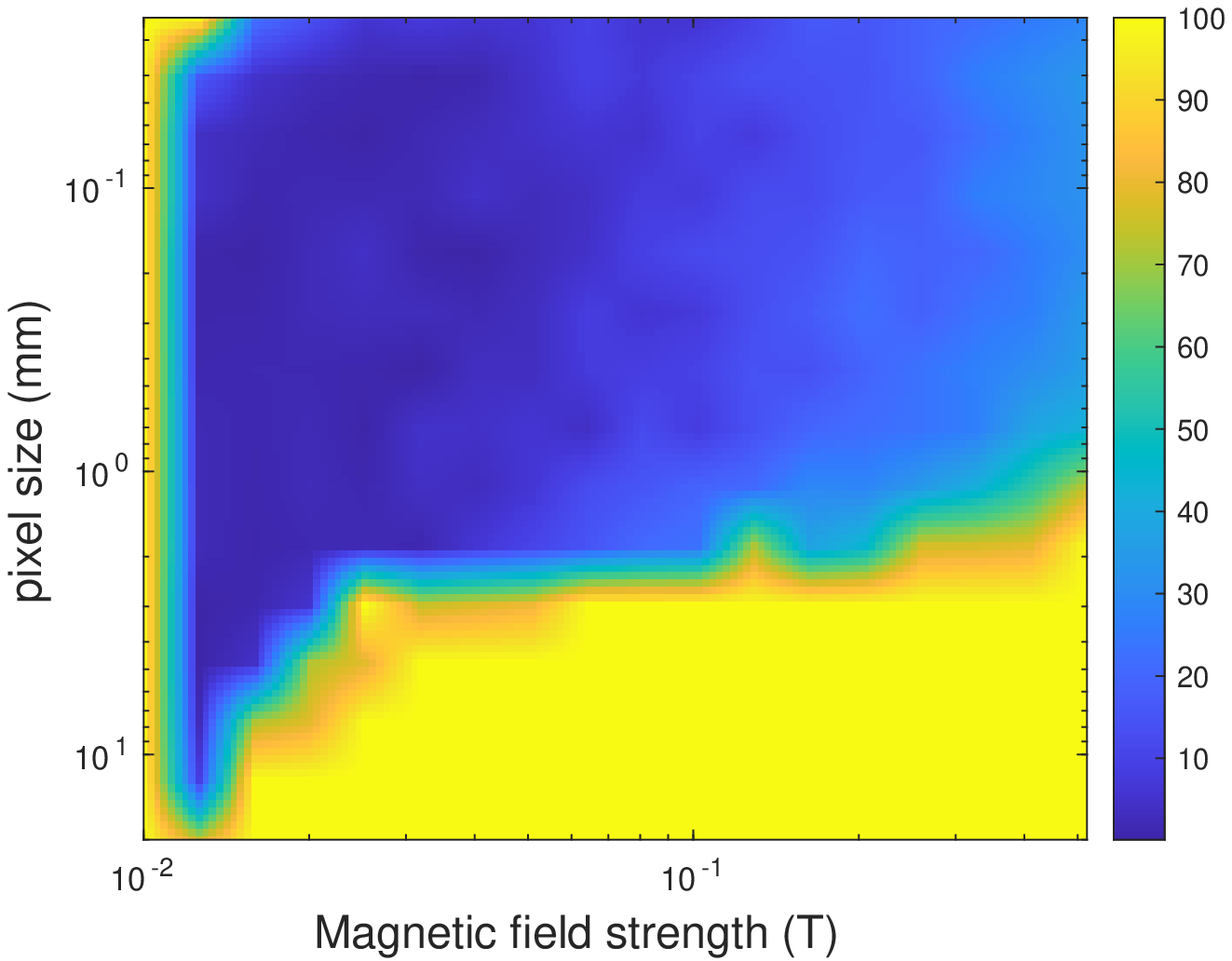}
 \subcaption{Density---$\kappa$-distribution}
 \label{density_kappa_B}
 \end{minipage}

 \begin{minipage}[t]{.45\linewidth}
 \centering
 \includegraphics[width=\textwidth]{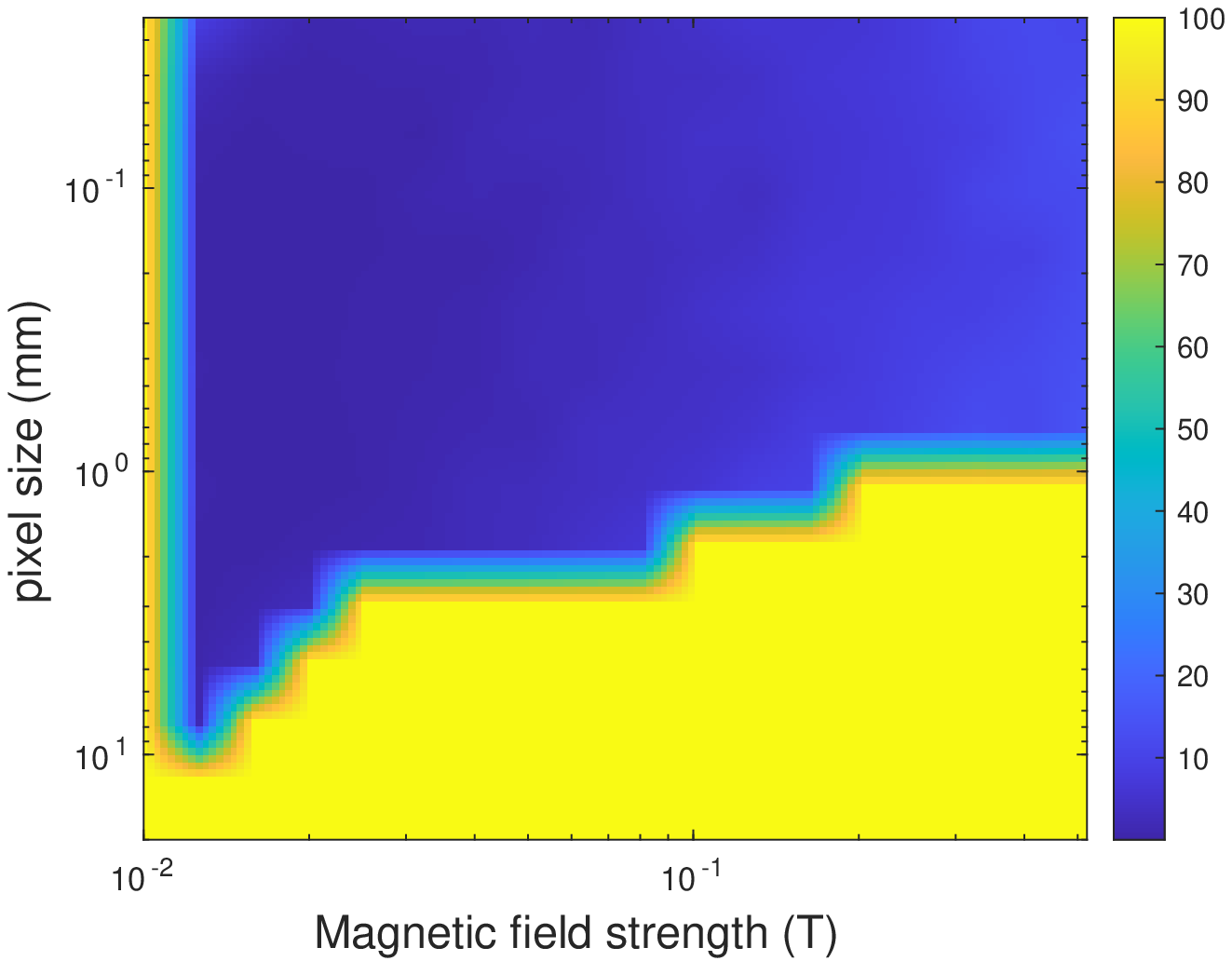}
 \subcaption{Thermal speed---Maxwellian distribution}
 \label{vt_maxwellian_B}
 \end{minipage}
 \hfill
 \vspace{0.2cm}
 \begin{minipage}[t]{.45\linewidth}
 \centering
 \includegraphics[width=\textwidth]{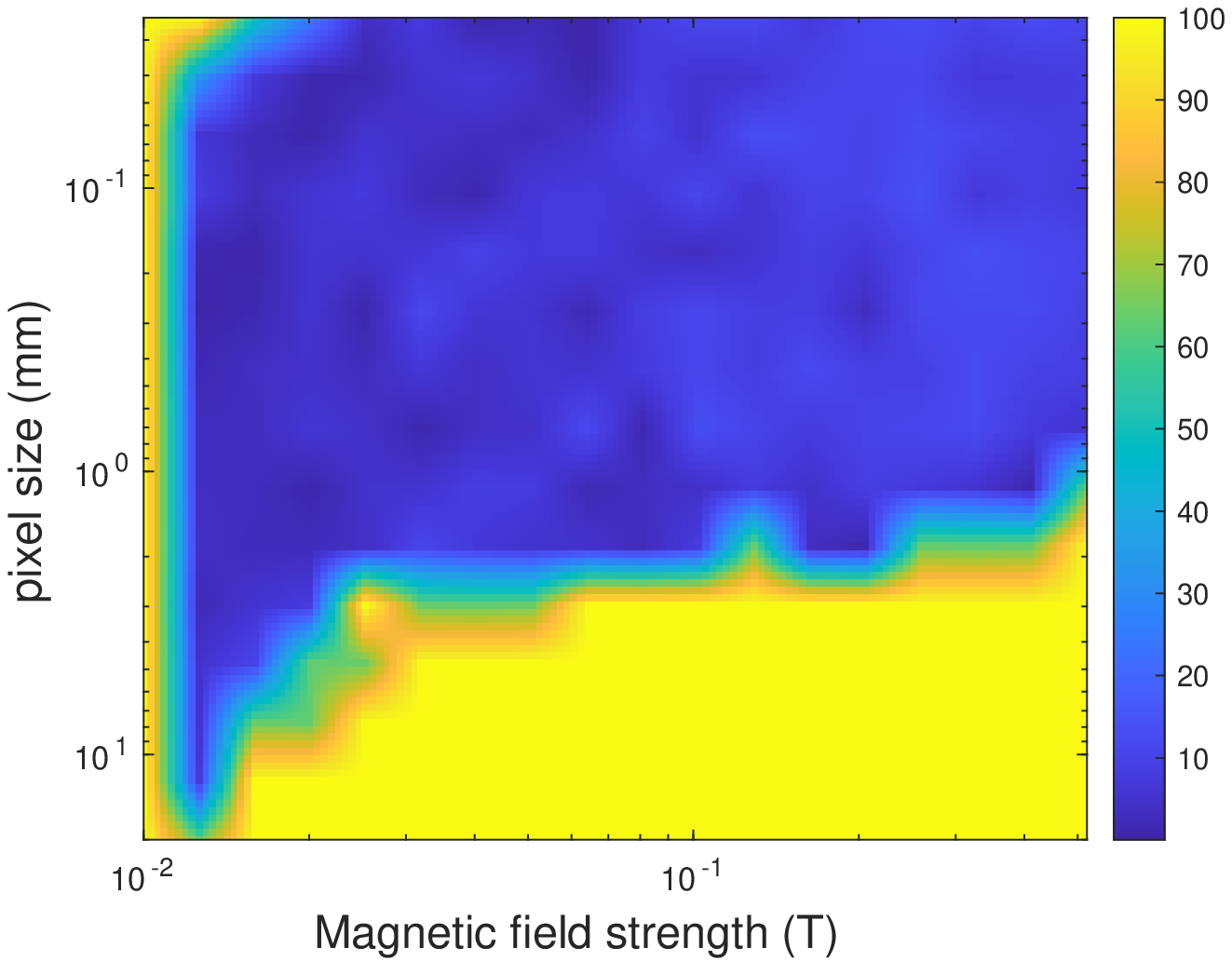}
 \subcaption{Thermal speed---$\kappa$-distribution}
 \label{vt_kappa_B}
 \end{minipage} 
	\hspace*{0.2cm}
	\begin{minipage}[t]{\linewidth}
 \centering
 \includegraphics[width=0.45\textwidth]{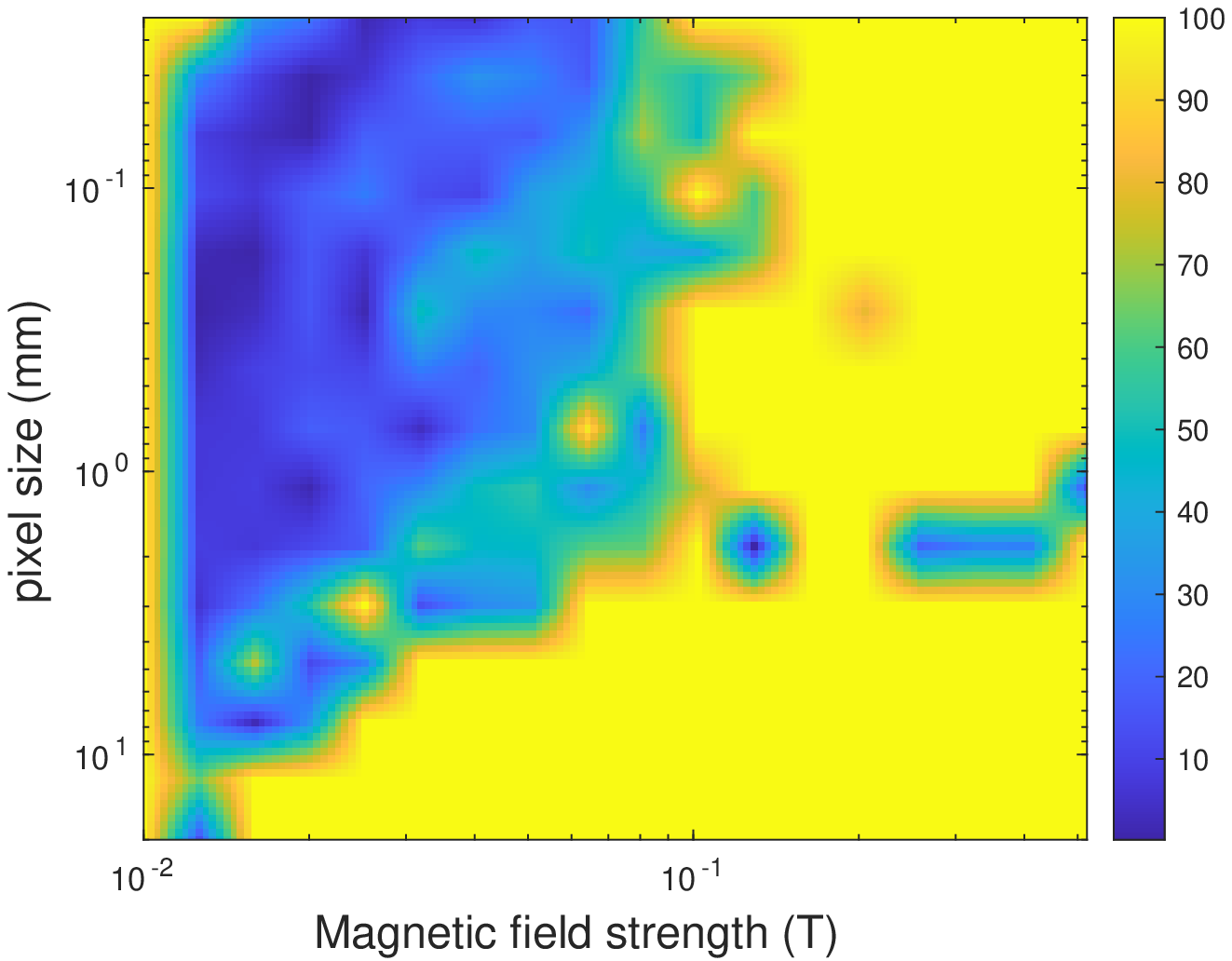}
 \subcaption{$\kappa$-parameter}
 \label{kappa_B}
 \end{minipage}
 \hfill \vspace{6pt} 
 \caption[Relative errors in percent made on measurements of an intermediate solar wind with magnetic field and pixel size.]{Relative errors in percent made on an intermediate solar wind with magnetic field and pixel size. Panels (\textbf{a}) through (\textbf{g}) show the resulting relative errors for all of the fit parameters of the Maxwellian and $\kappa$-distributions.}
 \label{figure:group_B}
\end{figure}


\section{Instrument Simulation}

The chosen geometry is summarized in Table \ref{table:instrument_characteristics} with an aperture radius of 1 mm and an acquisition time of 5 ms. With these parameters, we conduct ion tracing simulations with SIMION \cite{SIMION} to confirm the accurate and reliable functioning of MPA for both Maxwellian and $\kappa$-distributions. {In the following set of simulations, we evaluate whether MPA is able to retrieve the moments of Maxwellian and $\kappa$ VDFs as archetypical distributions to describe the solar wind \cite{kinetic_model_with_kappa, Verscharen} even though our instrument is able to resolve any other type of distribution with the same accuracy. The raw output of MPA corresponds to a list of speeds with their associated particle counts, so that the conversion to a distribution function only depends on the physical parameters of the instrument, as shown in Equations (\ref{equation:estimated})--(\ref{equation:true_counts}). Due to the linearity between $z_{\text{hit}}$ and $\|\textbf{v}_0\|$ in Equation (\ref{equation:z_hit}), MPA creates a direct one-to-one correspondence between the actual distribution and the measured spectrum.}

\subsection{SIMION Results for Protons}

\begin{figure}[H]
\centering
\includegraphics[scale=0.7]{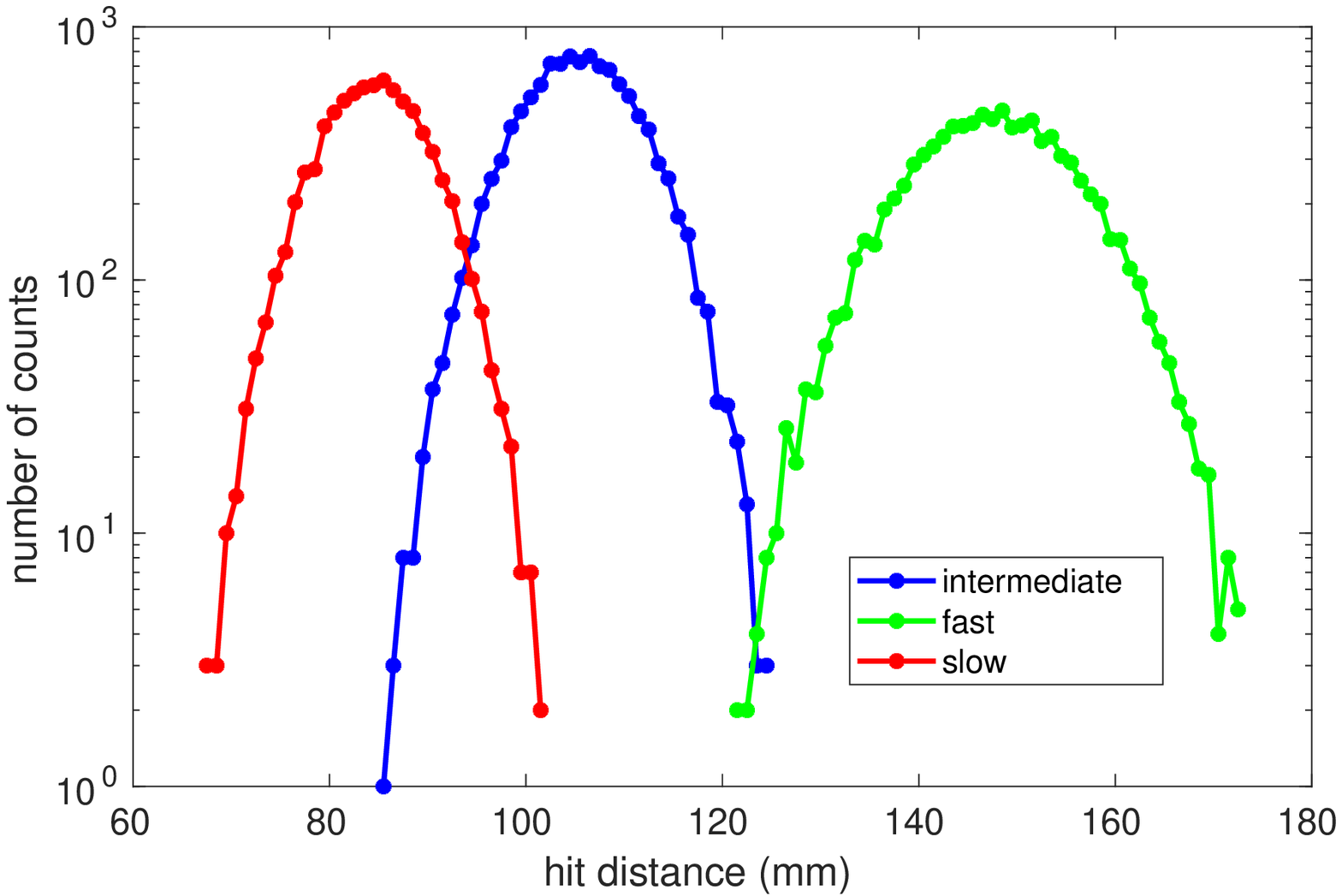}
\caption{Three different types of solar wind simulated with SIMION. We present the number of counts per pixel (1 mm wide) as a function of hit distance on the position-sensitive sensor along the $z$-axis in our instrument design. In all   three examples, the velocities of the simulated particles follow a Maxwellian distribution function.}
\label{figure:SIMION_maxwellian_counts}
\end{figure}

\begin{figure}[H]
\centering
\includegraphics[scale=0.7]{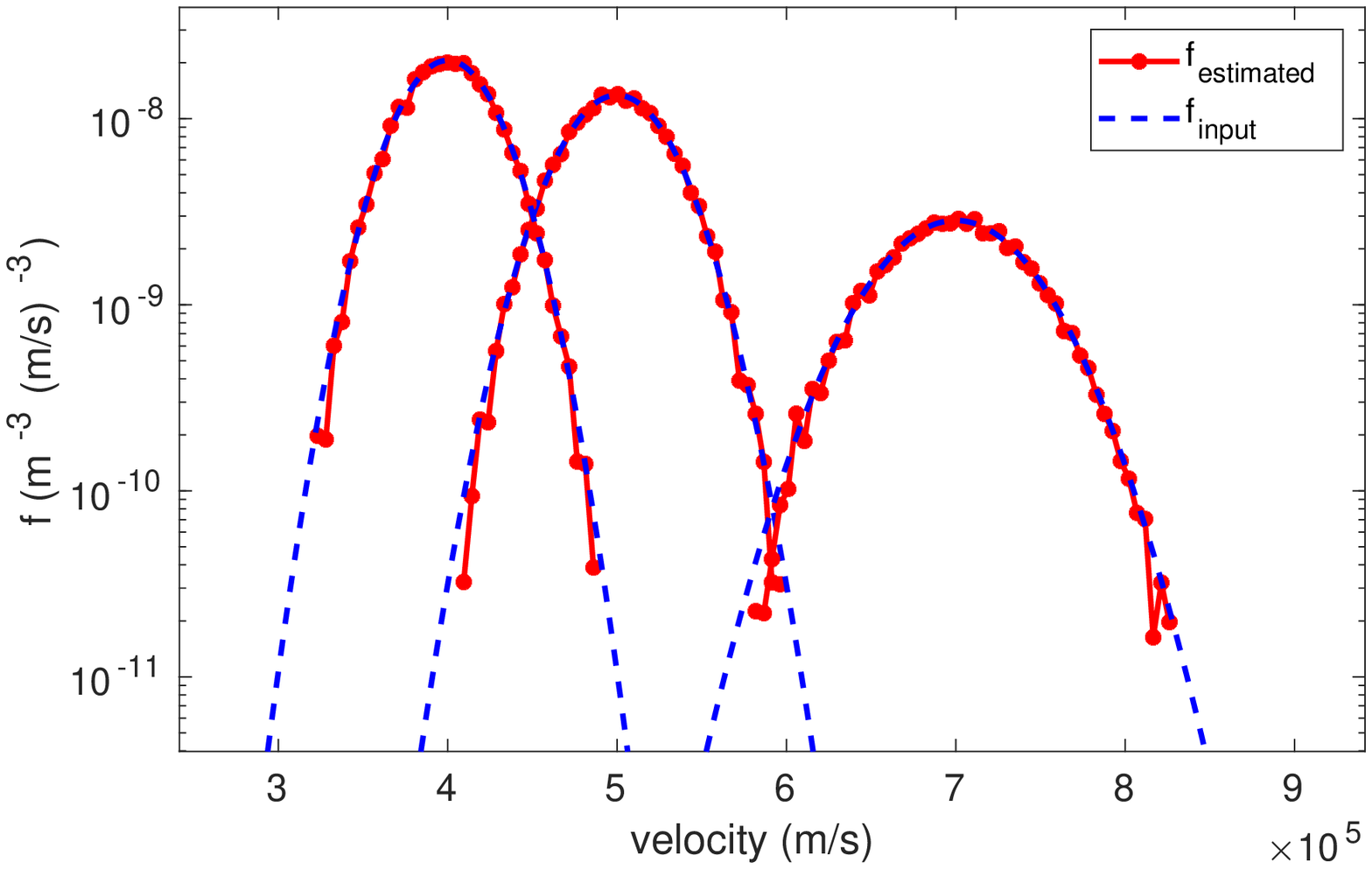}
\caption{Estimated (red) and model-input (blue) distribution functions. The estimation of the distribution function is based on the counting results shown in Figure \ref{figure:SIMION_maxwellian_counts}. In all   three examples, the velocities of the simulated particles follow a Maxwellian distribution function.}
\label{figure:SIMION_maxwellian_dist}
\end{figure}

\begin{figure}[H]
\centering
\includegraphics[scale=0.7]{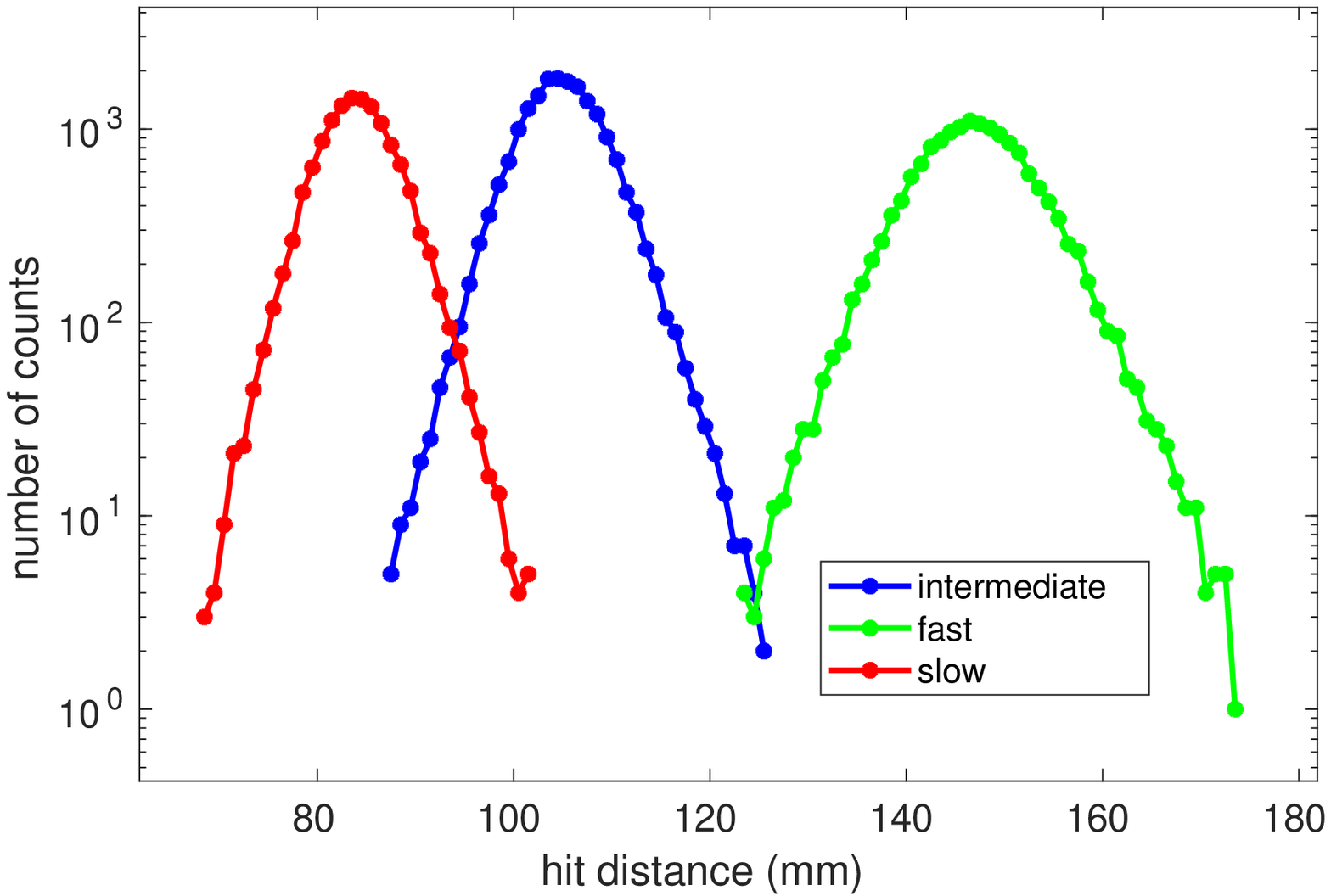}
\caption{Three different types of solar wind simulated with SIMION. We present the number of counts per pixel (1 mm wide) as a function of hit distance on the position-sensitive sensor along the $z$-axis in our instrument design. In all   three examples, the velocities of the simulated particles follow a $\kappa$-distribution function with $\kappa = 3$.}
\label{figure:SIMION_kappa_counts}
\end{figure}

\begin{figure}[H]
\centering
\includegraphics[scale=0.7]{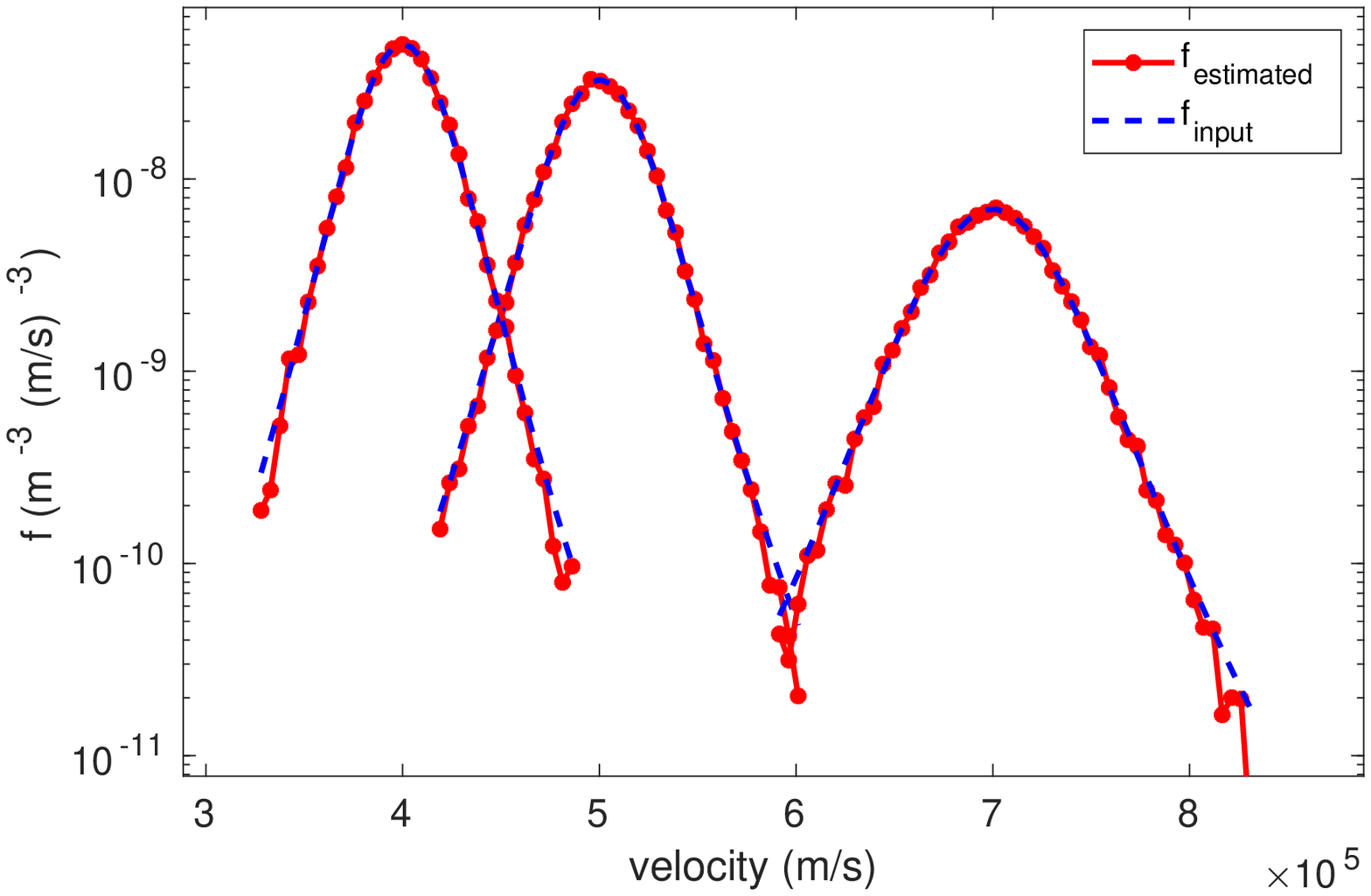}
\caption{Estimated (red) and model-input (blue) distribution functions. The estimation of the distribution function is based on the counting results shown in Figure \ref{figure:SIMION_kappa_counts}. In all   three examples, the velocities of the simulated particles follow a $\kappa$-distribution function with $\kappa = 3$.}
\label{figure:SIMION_kappa_dist}
\end{figure}

We simulate three different types of solar wind: slow, intermediate and fast (for parameters, \mbox{see Table \ref{table:sw_parameters}}). The SIMION results are presented in Figures \ref{figure:SIMION_maxwellian_counts} and \ref{figure:SIMION_maxwellian_dist} for Maxwellian distributions  and Figures \ref{figure:SIMION_kappa_counts} and \ref{figure:SIMION_kappa_dist} for $\kappa$-distributions. Figures \ref{figure:SIMION_maxwellian_counts} and \ref{figure:SIMION_kappa_counts} present the count numbers recorded on each pixel as a function of hit distance. We then use Equation (\ref{equation:estimated}) to obtain the associated VDFs and show the results in Figures \ref{figure:SIMION_maxwellian_dist} and \ref{figure:SIMION_kappa_dist}. The fit results of the estimated distribution functions are summarized in Tables \ref{table:result_fitting_maxwellian} and \ref{table:result_fitting_kappa}. MPA resolves the plasma parameters accurately over its 5 ms acquisition time. This~outcome supports the feasibility of MPA to give precise one-dimensional VDF measurements.

\begin{table}[H]
\caption{Slow, intermediate and fast solar wind input parameters for the Maxwellian and $\kappa$-distributions of speed used in SIMION simulations.}
\centering
\begin{tabular}{lcccc}
\toprule
\textbf{Solar Wind}	&	\textbf{Density (cm\boldmath{$^{-3}$})}	&	\textbf{Thermal Speed (m/s)}	&	\textbf{Bulk Speed (km/s)}	&	\textbf{$\kappa$}	\\
\midrule
Slow			&	5.5	&	36,341		&	400	&	3	\\
Intermediate	&	5	&	40,631		&	500	&	3	\\
Fast			&	3	&	57,460		&	700	&	3	\\
\bottomrule
\end{tabular}
\label{table:sw_parameters}
\end{table}
\unskip
\begin{table}[H]
\caption{Estimation of the plasma moments after fitting for our three Maxwellian distributions.}
\centering
\begin{tabular}{lccc}
\toprule
\textbf{Solar Wind}	&	\textbf{Density	(cm\boldmath{$^{-3}$})}		&	\textbf{Thermal Speed (m/s)}	&	\textbf{Bulk Speed (m/s)}	\\
\midrule
Slow		&	5.49 $\pm \,0.027$	&	36,416 $\pm \,68$	&	400,062 $\pm \,48$	\\
Intermediate		&	4.95 $\pm \,0.023$	&	40,559 $\pm \,72$	&	499,973 $\pm \,51$ 	\\
	Fast		&	2.94 $\pm \,0.015$ 	&	57,154 $\pm \,113$ 	&	700,029 $\pm \,80$	\\
\bottomrule
\end{tabular}
\label{table:result_fitting_maxwellian}
\end{table}
\unskip

\unskip

\begin{table}[H]
\caption{Estimation of the plasma moments after fitting for our three $\kappa$-distributions.}
\centering
\begin{tabular}{lcccc}
\toprule
\textbf{Solar Wind}	&	\textbf{Density	(cm\boldmath{$^{-3}$})}		&	\textbf{Thermal Speed (m/s)}	&	\textbf{Bulk Speed (m/s)}	&	\textbf{$\kappa$}	\\
\midrule
Slow		&	5.43 $\pm \,0.16$	&	34,538 $\pm \,1468$	&	399,981 $\pm \,68$	&	3.44 $\pm \,0.34$	\\
Intermediate		&	4.99 $\pm \,0.2$	&	39,447 $\pm \,2361$	&	499,994 $\pm \,101$	&	3.26 $\pm \,0.41$	\\
Fast		&	2.96 $\pm \,0.12$ 	&	56,492 $\pm \,3533$ 	&	699,963 $\pm \,138$	&	3.10 $\pm \,0.37$	\\
\bottomrule
\end{tabular}
\label{table:result_fitting_kappa}
\end{table}
\unskip

\unskip
\unskip
\vspace{-6pt}

\subsection{SIMION Results for Combined Proton and $\alpha$-Particle Measurements}\label{sec:simion}

$\alpha$-particles  are the second-most abundant ion species in the solar wind. We use a Maxwellian distribution to model the $\alpha$-particles. We use values from \citet{temperature_alpha} for the density and temperature of the $\alpha$-particles. Assuming an $\alpha$-particle-to-electron number-density ratio of 3.5\%, we~determine the Maxwellian distribution described in Table \ref{table:alpha_parameters}. We set the total density of positively charged particles to 5 cm$^{-3}$ and use a proton temperature of 100,000 K. The temperature ratio between protons and $\alpha$-particles is typically around 1.5. We neglect any relative drifts between $\alpha$-particles and protons, since these are small compared to their bulk speeds in the solar wind and would not affect our result significantly.

\begin{table}[H]
\caption{Properties of the $\alpha$-particle and proton input VDFs.}
\centering
\begin{tabular}{lccc}
\toprule
\textbf{Particle Species}	&	\textbf{Density (cm\boldmath{$^{-3}$})}	&	\textbf{Thermal Speed (m/s)}	&	\textbf{Bulk Speed (km/s)}		\\
\midrule
$\alpha$-particles	&	0.175	&	24,881	&	500	\\
Protons	&	4.825	&	40,631	&	500	\\
\bottomrule
\end{tabular}
\label{table:alpha_parameters}
\end{table}

\begin{figure}[H]
\centering
\includegraphics[scale=0.6]{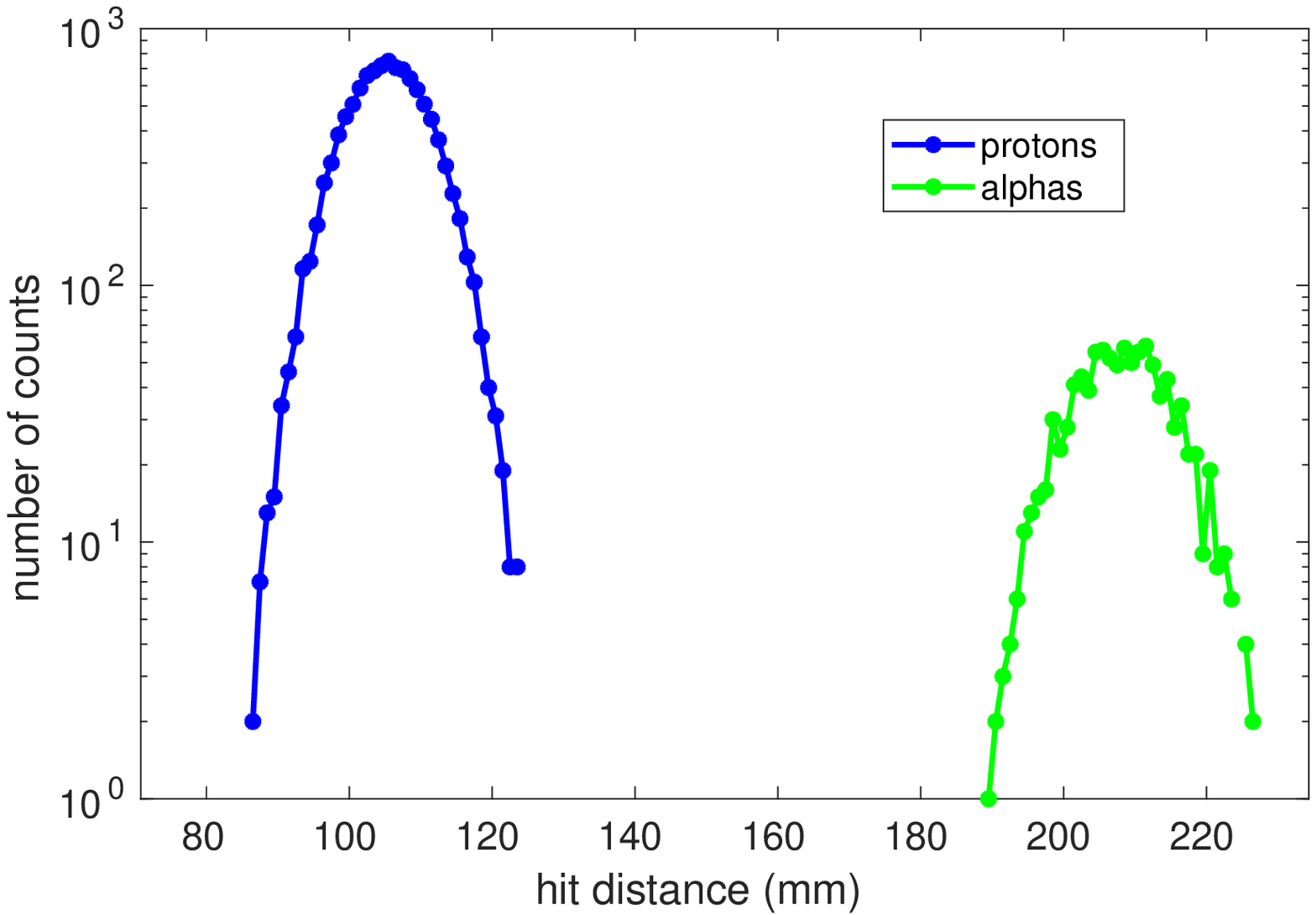}
\caption{Counts as a function of hit distance for $\alpha$-particles and protons.}
\label{figure:proton_alpha_counts}
\end{figure}

\begin{figure}[H]
\centering
\includegraphics[scale=0.6]{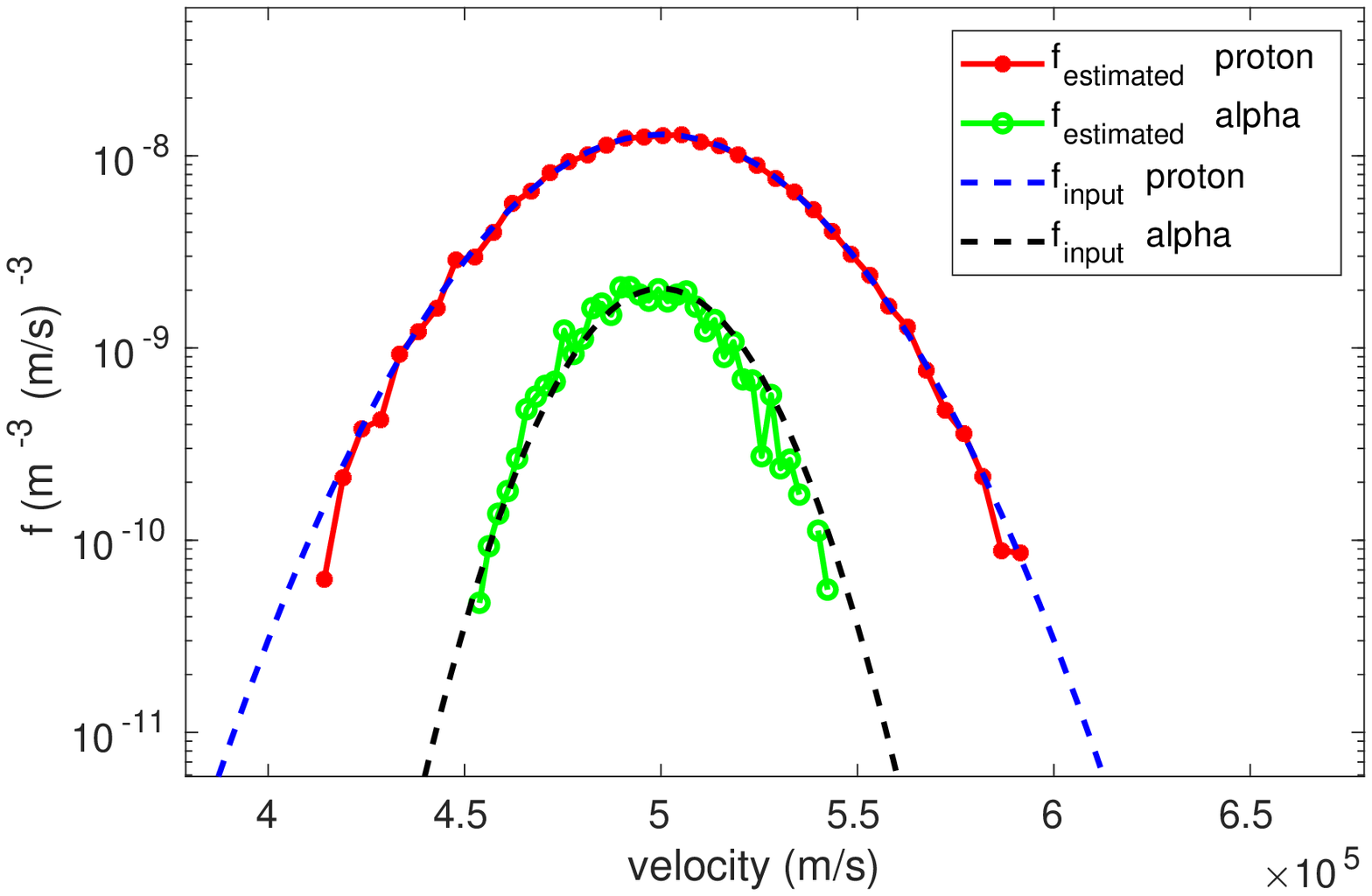}
\caption{Estimated (red and green) and model-input (blue and black) distribution functions. The~estimation of the distribution function is based on the counting results shown in Figure \ref{figure:proton_alpha_counts}.}
\label{figure:proton_alpha_distributions}
\end{figure}

Figures \ref{figure:proton_alpha_counts} and \ref{figure:proton_alpha_distributions} present our combined SIMION simulation results for protons and $\alpha$-particles. According to Figure \ref{figure:proton_alpha_distributions}, the $\alpha$-particles exhibit larger statistical fluctuations in the number of counts. However, as their thermal speed is higher, our instrument still provides us with a sufficient number of counts (around 80 at peak) to study this population. The fit results for the $\alpha$-particles are given in~Table\,\ref{table:alpha_fitting}.

\unskip
\unskip

\begin{table}[H]
\centering
\caption{Estimation of the plasma moments after a least square fitting for the simulated $\alpha$-particles and comparison with the corresponding input values.}
\begin{tabular}{lcc}
\toprule
\textbf{Parameter}	&	\textbf{Input Value}	&	\textbf{Output Value}	\\
\midrule
Density	&	0.175 cm$^{-3}$	&	0.160 cm$^{-3}$ $\pm$ 0.0043 cm$^{-3}$	\\
Thermal speed	&	24,881 m/s	&	24,151 m/s $\pm$ 249 m/s	\\
Bulk speed	&	500 km/s	&	496 km/s $\pm$ 176 m/s	\\
\bottomrule
\end{tabular}
\label{table:alpha_fitting}
\end{table}
\unskip

\section{Discussion and Conclusions}

Our instrument concept MPA aims at answering science questions on small-scale processes that are generally believed to be important for the solar-wind heating and acceleration. We study the ability of a magnetostatic plasma instrument to resolve proton and $\alpha$-particle VDFs with a high time resolution. We find that MPA can achieve this goal. MPA is based on the dependence of the gyro-radius of a charged particle travelling across a constant magnetic field on the particle velocity. The gyro-radius being proportional to the speed, by measuring the gyro-radius, we indirectly measure the particle speed. To obtain the best velocity resolution, we place our position-sensitive sensor such that it measures the diameter of the particle's circular motion. The velocity resolution of the instrument is then determined by the width of the pixels in the sensor plane, the aperture length and the value of the magnetic field strength. Whilst reducing the pixel size improves the velocity resolution, it deteriorates the counting statistics as  fewer particles hit each pixel. The key feature of MPA is that the Lorentz force naturally separates all incoming particles according to their speed. Thus, in one acquisition step, a full 1D cut of the VDF across all energies is obtained. This operating principle enables a high time resolution (5 ms). However, even in this instrument design, the time resolution is limited by counting statistics. To overcome inaccuracies due to counting statistics, we can enlarge the aperture,  by increasing  either the field of view or the aperture area, to the detriment of velocity resolution.

Our assumptions made for the sake of simplicity must be discussed and assessed in further detailed engineering work. A magnetic circuit simulation must be conducted to model the exact magnetic field between the two magnets. Shielding considerations are always important in the context of scientific space instrumentation, where magnetic cleanliness is crucial. A trade-off study between resolution and count rate will be necessary to decide on the best sensor, magnetic field strength and aperture geometry for our measurements. Moreover, the dependency of the count measurements on particle energy, mass, charge and hit angle must be addressed during calibration phases. Figure 1 of \citet{abs_detect_H} and Figure 3 of \citet{abs_detect_O} highlight the strong dependence of the detector efficiency on mass and charge of the analyzed particles. Consequently, it will be necessary to characterize the detector for ions such as $He^{2+}$, $C^{6+}$, $C^{5+}$, $O^{6+}$, $O^{7+}$, etc. found in the solar wind \cite{ion_comp, heavy_ion_comp}. However, since MPA separates ion species with identical inflow speeds by a distance proportional to $m/q$, a direct characterization of ion populations is possible.

Finally, MPA has a narrow field of view ($5^\circ\times5^\circ$), making 3D measurements impossible. Three-dimensional measurements of the solar wind   were achieved, for example, by \citet{Marsch} using Helios data between 0.3   and 1 au. These measurements have revealed that often temperature anisotropies occur with either $T_{\parallel}/T_{\perp} > 1$ or $T_{\parallel}/T_{\perp} < 1$ \cite{Marsch}, where $T_{\perp}$($T_{\parallel}$) is the temperature perpendicular (parallel) to the background magnetic field. These anisotropies are most pronounced at small heliocentric distances (around 0.3 au). Thus, 3D in-situ measurements are of interest to base and confirm analytical models of heating processes \cite{k_anisotropy, proton_diffusion} and when the VDF differs strongly from a Maxwellian. We note in this context that, with only one MPA-type instrument aboard a spacecraft, one would only be able to measure VDFs from one single look direction (notwithstanding the use of spacecraft spin to measure in multiple directions). Depending on the local field geometry at the time of the measurement, MPA samples cuts along varying directions relative to the local magnetic field. This leads to a lack of information on, for example, the instantaneous temperature anisotropy at the time of the measurement. However, the combination of multiple MPA units covering different look directions can resolve this shortcoming at the expense of mass and cost. Alternatively, we recommend the combination of MPA with a traditional ESA, so that MPA provides the high-resolution and high-cadence measurements in one look direction, while the ESA provides the 3D distribution at a lower resolution and cadence.

\subsection{Magnetic Field Design}

In all sections except Section \ref{sec:simion}, we assume a uniform field within the magnetic chamber ($B_0 = 0.1$\,T). The SIMION software, on the other hand, models a more realistic magnetic field by solving a Poisson equation for the magnetic field: $\textbf{B} = -\mu \nabla \Omega$, where $\Omega$ is the magnetic scalar potential and $\mu$ is the constant magnetic permeability. Thus, SIMION models more realistically the magnetic field generated by the two magnets in the absence of currents within the magnetic chamber. This~is sufficient for a first investigation at this point but not for a more detailed instrument design. The~path of the particles inside the instrument is governed by the instantaneous magnetic field vector at its current position. Furthermore, the magnetic circuit must be designed in detail to keep the field inside the gap as homogeneous and unidirectional as possible and the stray field as low as possible. This~work would benefit from the heritage of the MagEIS instrument in which case the magnet design was a key point of the success for the instrument \cite{MagEIS}. Indeed, in the Low and Medium units, the field's variations were kept below 0.5\% of the mean field value \cite{MagEIS} to ensure uniformity of the field. The stronger field inside the High chamber (0.48 T) of the MagEIS suite was realized with variations of around 5\% of the mean value.
Following the detailed design phase of the magnets for MPA, a file containing the established magnetic field vectors inside the chamber can be used in SIMION to simulate the real response of the instrument to different specific particle populations.

\subsection{Detectors and Readout Electronics}\label{sec:detector_tech}

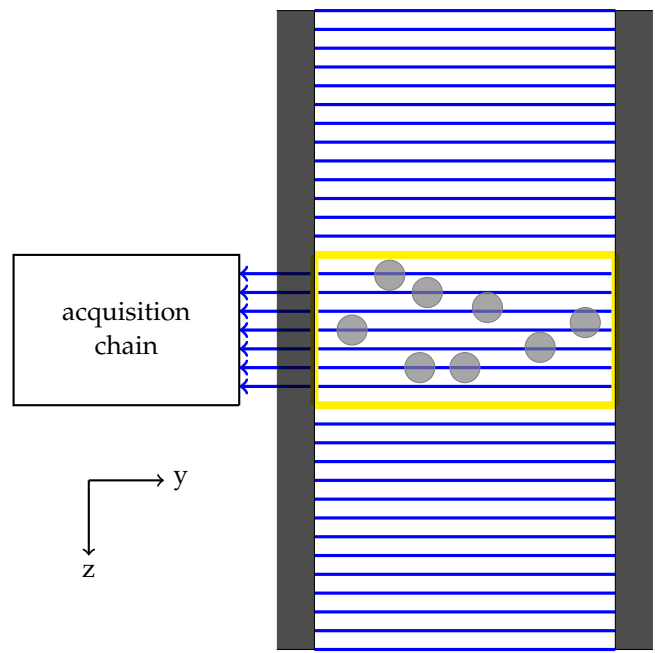
\begin{figure}[H]
\centering
\begin{tikzpicture}

\foreach \i in {-17,...,17}
{
	\draw[blue, very thick] (-2,\i/4) -- (2,\i/4);
}
\foreach \i in {-3,...,3}
{
	\draw[blue, very thick, ->] (-2,\i/4) -- (-3,\i/4);
}
\draw[black, thick] (-3,-1) -- (-3,1) -- (-6,1) -- (-6,-1) -- (-3,-1);
\node at (-4.5,0.2) {acquisition};
\node at (-4.5,-0.2) {chain};

\draw[yellow, line width=3pt] (-2,-1) -- (-2,1);
\draw[yellow, line width=3pt] (-2,1) -- (2,1);
\draw[yellow, line width=3pt] (2,1) -- (2,-1);
\draw[yellow, line width=3pt] (2,-1) -- (-2,-1);

\filldraw[gray, fill opacity=0.7] (-0.5,0.5) circle (0.2);
\filldraw[gray, fill opacity=0.7] (0,-0.5) circle (0.2);
\filldraw[gray, fill opacity=0.7] (0.3,0.3) circle (0.2);
\filldraw[gray, fill opacity=0.7] (-0.6,-0.5) circle (0.2);
\filldraw[gray, fill opacity=0.7] (-1,0.73) circle (0.2);
\filldraw[gray, fill opacity=0.7] (1,-0.23) circle (0.2);
\filldraw[gray, fill opacity=0.7] (1.6,0.1) circle (0.2);
\filldraw[gray, fill opacity=0.7] (-1.5,-0.0) circle (0.2);

\filldraw[black, fill opacity=0.7] (-2.5,-17/4) -- (-2,-17/4) -- (-2,17/4) -- (-2.5,17/4);
\filldraw[black, fill opacity=0.7] (2.5,-17/4) -- (2,-17/4) -- (2,17/4) -- (2.5,17/4);

\draw[black, thick, ->] (-5,-2) -- (-5,-3);
\node[below] at (-5,-3) {z};
\draw[black, thick, ->] (-5,-2) -- (-4,-2);
\node[right] at (-4,-2) {y};
	
\end{tikzpicture}
\caption{Position-sensitive anodes and a representative pixel.   The anodes are represented by the blue lines. They consist of conductive wires that detect the current generated by the electron beams (gray circles) and determine the vertical $z$ position of the beam based on the knowledge of which wire has been hit. A pixel is defined as either a single anode or a collection of multiple anode wires and is represented here as a yellow rectangle. The two permanent magnets, which delimit the magnetic chamber, are represented by the dark rectangles at both sides of the pixels.}
\label{schematic:anode}
\end{figure}

To record the hit position of protons and $\alpha$-particles, our instrument requires a position-sensitive sensor plane. This can  be achieved by  the use of either a micro-channel plate (MCP) or a channel electron multiplier (CEM). Moreover, we explicitly require a high time resolution while maintaining high count numbers. This can raise a concern regarding MCPs as their saturation count rate is around 1 MHz, depending on the detecting surface area. On the other hand, CEMs present higher count rates independent of the hit area but may induce a lower velocity resolution due to their size. \citet{THOR} use 32 CEMs to achieve a 1.5$^\circ$ resolution in their high-cadence THOR instrument design. For MPA, the~sensor has a rectangular shape (strip) and pixels are organized along the $z$ axis. Figure \ref{schematic:anode} shows an anode array in the case of an MCP detector. 

While we do not show the MCP itself in this figure, we represent the MCP-produced electron clouds as gray circles. An acquisition chain is responsible for recording every hit and its position along the $z$ axis. A pixel (yellow rectangle in Figure \ref{schematic:anode}) can correspond to a unique anode wire or resistive strip, or a combination of neighboring wires/strips, depending on the trade-off between velocity resolution required for a particular measurement and counting statistics (the summation of counts from multiple wires/strips defines the number of counts in a pixel if multiple wires/strips are combined). We recommend that   readout electronics composed of amplifiers and discriminators  are multiplexed to reduce the number of components required.

\vspace{6pt} 


\authorcontributions{Conceptualization, B.C. and G.N.; methodology, B.C., G.N. and D.V.; software, B.C.; validation, G.N. and D.V.; formal analysis, B.C.; investigation, B.C.; writing---original draft preparation, B.C.; writing---review and editing, B.C., G.N. and D.V.; visualization, B.C.; supervision, G.N. and D.V.; project~administration, G.N. and D.V.; and funding acquisition, D.V. All authors have read and agreed to the published version of the manuscript.}

\funding{This research was partly funded by STFC Consolidated Grant ST/S000240/1 and STFC Ernest Rutherford Fellowship ST/P003826/1.}

\acknowledgments{The authors thank Robert T. Wicks for helpful discussions. The code written for the simulations in this publication is publicly available at \url{https://github.com/c1-94/MPA}.}

\conflictsofinterest{The authors declare no conflict of interest. The funders had no role in the design of the study; in the collection, analyses, or interpretation of data; in the writing of the manuscript, or in the decision to publish the results.} 

%
%



\reftitle{References}





\end{document}